\documentclass[preprint]{aastex}

\slugcomment{22Mar2005 - Accepted for publication in {\sl The
    Astronomical Journal\/}}

\shorttitle{High Proper Motion Stars in USNO-B1}
\shortauthors{S. Levine}

\received{2005 February 1}
\begin{document}

\title{Stars in the USNO-B1 Catalog with Proper Motions \\    
  Between 1.0 and 5.0 arcseconds per year}

\author{Stephen E. Levine}
\affil{U. S. Naval Observatory, Flagstaff Station, 10391 West Naval
  Observatory Road, Flagstaff, AZ 86001-8521, USA}
\email{sel@nofs.navy.mil}

\begin{abstract}

This paper examines a subset of objects from the USNO-B1 catalogue
with listed proper motions between 1.0 and 5.0 arcseconds per year.
We look at the degree of contamination within this range of proper
motions, and point out the major sources of spurious high proper
motion objects.  Roughly 0.1\% of the objects in the USNO-B1 catalogue
with listed motions between 1.0 and 5.0 arcseconds per year are real.
Comparison with the revised version of Luyten's Half Second catalogue
indicates that USNO-B1 is only about 47\% complete for stars in this
range.  Preliminary studies indicate that there may be a dip in
completeness in USNO-B1 for objects with motions near 0.1 arcseconds
per year.  We also present two new stars with motions between 1.0 and
5.0 arcseconds per year, 36 new stars with confirmed motions between
0.1 and 1.0 arcseconds per year, several new common proper motion
pairs, and the recovery of LHS~237a (VBs3).

\end{abstract}


\section{Introduction}

Can the USNO-B1 catalogue \citep{metal03} be used to find previously
unknown objects with large proper motions?  Our motivation for
examining the set of objects with large proper motions that are in the
USNO-B1 catalogue is two-fold.  First, we are looking to see if we
have found any objects with large proper motions that were missed in
previous surveys; these could well be interesting in their own right,
and for studies of the local neighborhood.  Second, we would like to
understand how well the motion finding algorithm used in the
construction of USNO-B1 worked, and how contaminated is the high
motion sample.

Because objects with large proper motions tend to be relatively
nearby, they are often intrinsically interesting astronomically.
Given this consideration, we would like to know what fraction of
objects in USNO-B1 with listed large motions are actually real, and
how we can go about selecting clean samples of objects with large
motions.  In the construction of USNO-B1, \citet{metal03} erred on the
side of retaining dubious objects.  Some of the lessons learned here
about cleaning up sub-samples should be readily applicable to many
other subsets of objects in USNO-B1.

\citet{g03} has already examined in some detail how well USNO-B1 has
done in finding previously known objects that are in the revised
version of Luyten's Two Tenths Catalogue \citep[][hereafter
NLTT]{nltt,gs03,sg03} (which contains star with $\mu > 0.180 \, {\rm
arcsec \, yr^{-1}}$).  Only 6\% of the NLTT stars are missing from
USNO-B1, and an additional 4\% have what they term large errors
\citep{g03}. Hence, their conclusion was that USNO-B1 is roughly 90\%
complete, with incompleteness rising at both the brighter and fainter
ends.  They also find that incompleteness increases at larger proper
motions (it is roughly 30\% at $\mu = 1 \, {\rm arcsec \, yr^{-1}}$),
and near the galactic plane.  The proper motion values given in
USNO-B1 are generally in agreement with those in the NLTT.

The aim here is complementary.  We are looking at the contents of the
catalogue, and trying to assess what fraction of the moving objects in
the high motion portion of the catalogue are in fact real objects.  We
also wish to know if the derived motions are reasonable.  For the
entries in the catalogue that correspond to non-real objects, we hope
to gain some understanding of their characteristics, and hence learn
how to exclude them in the future.

The USNO-B1 catalogue is an all-sky catalogue that has been compiled
from digitizations of 7,435 Schmidt plates taken over the last 50
years (see Table~\ref{tbl-plates} for a summary of the plate material
that was used).  Every point on the sky is covered at several epochs and at
several wavelengths, making it possible to construct a catalogue that
includes positions, proper motions, optical colors, star/non-star
discriminators and the appropriate uncertainties \citep{metal03}.
This catalogue is the natural successor to the USNO-A series of
catalogues \citep{metal96,metal98}, and should fix a number of
problems associated with them.  Because of the nature of the plates,
the images (and hence the catalogue) are best for fainter objects (in
the magnitude range $V = 14$ to $22$).  The Tycho-2 catalogue
\citep{hetal00} was copied in for completeness at brighter magnitudes
(Tycho-2 is complete at the 99\% level down to $V_T=11$).  In regions where
confusion is not the limiting factor, the catalogue is complete to
photographic magnitudes ${\rm B_{phot} \approx 21}$ and ${\rm R_{phot}
\approx 20}$ \citep{metal03,metal04}.

It is prudent to note that USNO-B1 is an inclusive catalogue,
by which we mean that in the construction of USNO-B1, \citet{metal03}
erred on the side of including all possible objects, real and false.
The aim was to avoid removing real objects during the assembly of the
catalogue, and to give users some flexibility in designing their own
selection algorithms.  One of the results however is that some fraction
of the objects in the catalogue are either contaminated, or completely
false, and it is desirable to avoid selecting these entries.

One of the key improvements of this catalogue with respect to its
predecessors is the determination of proper motions for all objects in
the catalogue.  Proper motions provide important information about the
motions of objects, and about the structure of our Galaxy. In
addition, the proper motion can be a very useful discriminant when
trying to find objects meeting specific criteria (e.g. objects that
are close to us or those in the halo often have large apparent proper
motions).  To see this, one needs only to look at how fruitful studies
have been of the objects in the catalogues of high proper motion stars
of \citet[][hereafter LHS and NLTT respectively, where LHS is the
Luyten Half Second Catalogue]{lhs,nltt} and \citet{gic71,gic78}
(e.g. proper motion information has aided in the selection of nearby
objects for study), and also how much has been learned about things
like the structure of the local neighborhood in the Galaxy from the
Hipparcos \citep{hip} and Tycho-2 \citep{hetal00} catalogues
\citep[e.g.][]{db98,od03}.

The proper motions in the USNO-B1 catalogue have some known
idiosyncrasies.  Among these are that the motions given are strictly
relative proper motions, since least squares has set the mean motion
for stars of roughly $18^{\rm th}\, {\rm magnitude}$ to zero on a
field by field basis.  The component of solar motion relative to this
zero point is small when compared with the motions we are interested
in here.

Both \citet{metal04} and \citet{gk04} have produced improved proper
motion catalogues for the region contained in the intersection of the Sloan
Digital Sky Survey Data Release 1 \citep[][hereafter SDSS DR1]{aetal03} 
with USNO-B1.  In both cases, the contamination problem has largely been
dealt with by using SDSS DR1 data as truth, and re-calibrating the
overlapping region of USNO-B1.  In addition, both catalogues use 
sources external to our galaxy like galaxies \citep{metal04} or quasars
\citep{gk04} found in the SDSS sample to put the revised proper motions
onto an absolute scale.   \citet{metal04} note, and we concur, that
as the number of detections of an object in USNO-B1 decrease from the
maximum possible of 5, the likelihood that the object is contaminated
or totally false increases dramatically \citep[see][Fig. 11]{metal04}.

Section 2 explains briefly the moving object detection algorithm used
in the construction of the USNO-B1 catalogue.  Sections 3 and 4
discuss how we went about finding fast moving objects in the
catalogue, what portion of the listed objects are real, and what
additional objects we found along the way.  Notes about specific
objects are given in section 5.  Section 6 discusses briefly a
comparison of the high proper motion samples in LHS and USNO-B1.  The
paper concludes with a bit of discussion.

\section{Moving Object Detection Algorithm}

In the construction of USNO-B1, finding objects with large proper
motions was handled as a special case of measuring proper motions for
all objects \citep[the discussion here closely follows][which should
be referred to for more complete details]{metal03}.  The search for
moving objects can be broken down into two parts: the first part was to
find objects that do not move, or move only a little bit, and then the
second was to look for objects with large motions.

In the catalogue construction, the sky was broken up into rings of
$0\fdg1$ width in declination.  Each ring initially contained the complete
set of detections from all of the plates (at all epochs) that
intersected that ring.  Though magnitudes in USNO-B1 are referred to
as being of first or second epoch, in reality, the second epoch can
cover a wide range of time: for example, for POSS-II, the 3 ``second
epoch'' plates could cover as much as 10 or 15 years of time.  A three
arcsecond aperture was moved through each ring.  The cases where only a
single detection fell within the aperture were ignored at this point.
If detections from one or more first epoch surveys and one or more
second epoch surveys were within this aperture, then these detections
were matched up as an object and those detections were removed from
the lists.  This should have matched up detections for objects that
move less than about $60 \, {\rm mas\, yr^{-1}}$.

Only those detections which were not matched up under the slow motion
search radius, were passed on to the high proper motion search
routine.  For this step, when searching a band in declination, the two
adjoining bands were also included so that in fact the search regions
were $0\fdg3$ in width, in steps of $0\fdg1$ in declination.  The
search aperture was expanded to 30 arcseconds.  Within the aperture,
all combinations of second epoch detections were fit for linear
motion.  If a fit was significant, then the motion was extrapolated
back to the first epoch surveys.  A search around the predicted point
was done, using a search radius that scaled with the size of the
extrapolated error ellipse.  All possible combinations with first
epoch observations were followed up.  If the best fit had a standard
deviation less than than 0.4 arcsec in the tangent-plane coordinates,
and a motion less than 10, 3 or 1 ${\rm arcsec \, yr^{-1}}$ for 5, 4
or 3 survey detections respectively, then the object was considered
matched and the detections were removed from the lists.

After the explicit search for high motion objects, a last effort was made
to match up any remaining objects.  A search aperture of 20 arcseconds
was used, and all combinations of 5, 4, 3, and 2 survey detection
objects were examined.  The first groups of observations with a standard
deviation of less than 5 arcseconds in both tangent plane coordinates
were considered matched, called an object, and removed from the
detection lists.  Only a small fraction of the objects in USNO-B1 came
from this step in the processing.  Presumably, many matches made at this
stage will look like high motion objects.

It is useful to point out that no use was made in the matching process
of magnitude data, or star/non-star separator information.  After the
object matching was done, duplicate objects were removed
\citep{metal03}.

\section{Selecting Real High Motion Objects}

Retrieving real objects from USNO-B1 with large proper motions is
not quite as simple a task as just asking for all of the catalogue
objects with motions in a given range.  To demonstrate this, we chose
to search for previously unknown objects with proper motions ($\mu$)
between 1.0 and 5.0 ${\rm arcsec \, yr^{-1}}$.

We began by requesting all objects in USNO-B1 with $1.0 \le \mu_{\rm
total} \le 5.0 \, {\rm arcsec \, yr^{-1}}$.  This netted a total of
187,134 objects (see Fig.~\ref{fig-one}(a) and (d)).  We then applied
some basic sanity checks.  We required each object to be detected on 4
or 5 of the 5 possible surveys ($N_{FitsPts} \ge 4$) used in the
construction of USNO-B1.  This reduced the number of objects to
186,554.  This also had the effect of removing all of the Tycho-2
stars that were added in, since the number of survey detections for
these stars was set to 0 in the catalogue.  Then, we required that
they have position errors less than 0.999 arcsec in both RA and Dec
($\sigma_\alpha < 999 \, {\rm mas}$ and $\sigma_\delta < 999 \, {\rm
mas}$).  This decreased the number of objects to be considered to
11,019 (see Fig.~\ref{fig-one}(b) and (e)).
Further limiting the sample by applying limits to the second epoch
photographic R magnitude ($0 \le {\rm R_2} \le 18.0$), brought the
total down to 3,348 objects (see Fig.~\ref{fig-one}(c) and (f)).  Not
too surprisingly, after the application of the basic sanity checks,
most of the potential objects lie near the galactic plane, and near
the celestial poles (where the plate overlap regions grow larger).

This remaining sample of 3,348 objects was next examined by eye.  The
catalogue data around each potential high motion object were plotted
and a decision was made as to the likelihood that this was a real
object with a large proper motion.  Typical things to look for and
select against were diffraction spikes and other artifacts caused by
bright stars (see Fig.~\ref{fig-diff}(center)) and extended objects.
These all tended to produce groups of objects in the catalogue that
are closely clumped, or show obvious large scale structure (like the
linear arms of the diffraction spikes, or the arcs of the halos of
bright star ghost images).  951 objects passed this somewhat
subjective test.  Of the other 2,397 potential objects, only 7
turned out to be real, already known objects that we failed to
recognize.  672 objects
were rejected as being caused by proximity to a bright star or a
bright star's diffraction spikes.  1,725 objects
were mis-identifications and mis-matches caused by a
variety of forms of confusion (dense field, extended object(s) that
created multiple detections, etc).  All 7 missed real objects fell
into the latter category.

Of the 951 objects reaching this stage, 177 are stars that were already
flagged in USNO-B1 as being known proper motion objects, leaving us
with 774 candidates to check.  The objects that were left were
presumed to be decent candidates for being new high proper motion stars.
Images from several epochs were extracted from the USNO Image and
Catalogue Archive\footnote{http://www.nofs.navy.mil/data/fchpix/} and
the images of each potential object were again looked at by eye.  Of
these, 741 objects turned out to be confused, diffraction spikes, near
a bright star, or extended objects,
leaving only 33 real, moving stars.  To this, we added one more star
found during preliminary testing.

The 34 real, unflagged moving objects were then checked against
catalogues of known objects.  Nine were found to be LHS objects that
somehow did not get flagged in the construction of the catalogue
(including LHS 237a, which we have recovered, and about which we have
more to say below).  (As a side note, it is worth pointing out that
just by counting the number of objects in the USNO-B1 catalogue with
the ``known proper motion star'' bit set shows that a fair fraction of
these objects that were already known did not get flagged properly in
the catalogue.)  Seven turned out to have been found recently by
\citet[hereafter LSR]{lsr02}, and three more or less simultaneously by
us and \citet{l05}.  One was a brown dwarf found by the DENIS survey
\citep{detal01}(DENIS-P J104814.7-395606.1). One was a halo dwarf
found by \citet{oetal01} (WD0205-053).  Two other objects were found
recently by \citet{hetal04} using SuperCOSMOS (SCR0342-6407 and
SCR2012-5956).  Three were found by \citet{petal03,petal04}(LEHPM
4051, 3861, 4466), one by \citet{rrsi02}(APMPM J1957-4216), and one
object was recently found by \citet{dhc05}(SIPS0052-6201) (see
Table~\ref{tbl-known}).

Of the 6 remaining objects, four are moving objects with incorrect
proper motions in USNO-B1; the objects are moving, just more slowly
than the catalogue indicates.  One of these objects is NLTT 9526, and
the other three appear to be new (see Table~\ref{tbl-ltone}).  The
final 2 objects are (to the best of our knowledge) new, high proper
motion stars (see Table~\ref{tbl-new}).  The 30 objects with proper
motions larger than 1 arcseconds are plotted in Fig.~\ref{fig-thr}(a)
and (b) as filled triangles for the new objects and as 3--pointed
stars for previously known, but unflagged objects.  The 4 with proper
motions less than 1 arcsecond are plotted as filled hexagons for the 3
new ones, and as a 6--pointed star for the one already known.  Finder
charts showing the plate material for the 2 new high proper motion
objects are given in Figures~\ref{fig-hm0484} and \ref{fig-hm0867}.
In all, 6\%, or 213, of the 3,348 objects are stars with motions
between 1 and 5 ${\rm arcsec \, yr^{-1}}$.

In the 774 fields that were examined by eye, over 71 of them had a
total of 82 other objects with apparent proper motions.  These are
discussed in more detail below in section 4 on Serendipitous objects.

\subsection{How to speed up the winnowing process}

Did we learn anything from the winnowing process described above that
would help us to generate more easily clean(er) sub-samples out of
USNO-B1?  Without using information from outside of USNO-B1, the
following simple things can be done quickly to help preselect for
objects that are likely to be real.

Specify that objects have positional errors less than 999~mas in both
RA and Dec.  This reduced the sample by over an order of magnitude
from $\sim 10^5$ objects to $~10^4$ objects.  In Fig.~\ref{fig-hisigs}
are plotted the distributions of position and motion errors of the
3,348 objects examined by eye.  We see that almost all the real
objects have a total positional error less than 350 mas.  The false
objects have a much wider distribution.  An optimal cut is probably
closer to 350 to 500~mas.

The next simple cut that we can apply is to the proper motion error
and is based on Fig.~\ref{fig-hisigs} (right column).  This shows that
most of the real high motion objects have a proper motion error less
than $12 \, {\rm mas \, yr^{-1}}$, while the false objects have a
larger secondary hump at $30 \, {\rm mas \, yr^{-1}}$.

Another thing that can be done is to insist that objects be detected
on at least 4 surveys.  The benefit from this one is a little less
clear.  For the objects with large proper motions, insisting on
detection on 4 or 5 surveys out of a possible 5 only removed 580
objects out of 187,134 (or 0.3\% of the total) in the original search.
It is also a reasonable to presume that the position and proper motion
errors will be anti-correlated with the number of detections.  On the
other hand, this is a very quick and simple culling criterion to
implement, as the number of detections is carried as an integer in
each object's catalogue record.  In addition, for objects with lower
motions, there will be more catalogue objects with 3 or 2 detections,
hence this will likely be more useful for searches of things other
than the high proper motion objects.

It is also instructive to be aware of where in the sky you are
looking.  In Fig.~\ref{fig-one}(a), there is a change across the line
of $\delta \approx -33^\circ$ that is largely due to the difference in
the number of first epoch plates, and the epoch difference between the
first and second epochs.  North of this line, the first epoch Palomar
Observatory Sky Survey (hereafter POSS-I) provides two plates at a
mean epoch near 1950. South of that line, the first epoch is a single
red plate with a mean epoch around 1980. There is a much shorter
southern temporal baseline, and there is one fewer plate per field.
Internal tests done during the construction of USNO-B1 showed that
each additional plate dramatically reduced the false positive rate
when looking for high motion objects (D. Monet, private
communication).  This is not particularly surprising, as the motion is
presumed to be almost linear, and it becomes increasingly unlikely
that N random points will be nearly co-linear as N increases.

We can now apply magnitude related criteria.  This was not done in the
original search, but is a simple test that enforces an additional
degree of consistency upon the data.  Requiring that the difference
between the first epoch red magnitude ($R_1$) and the second epoch red
magnitude ($R_2$) be less than 0.5 (or 1.0) magnitudes helps to
exclude improperly matched detections.

We re-did the search for high proper motion objects, this time
applying all the criteria listed in this section, EXCEPT that for the
number of plates criterion, we allowed either 0, 4 or 5 plates to be
accepted (this meant we included the Tycho-2 stars, which are the only
ones in the catalogue with a value of 0 plates set).  After applying
cuts based solely on the position and motion errors the sample size is
reduced to 8,576, and includes 196 of 207 known and flagged high
proper motion objects (not including Tycho-2 objects).  Once magnitude
related cuts are applied, the data volume is reduced to 1,478 objects
where $|R_1-R_2| < 0.5 \, {\rm mag}$ (or 2,556 for a magnitude
difference of 1), including 137 (168) of the already known high proper
motion objects.  Finally, when we limited our list to $R_2$ brighter
than magnitude 18 (the same de facto restriction we used in generating
our original list), that left us with 688 (or 1,090) objects out of an
original 187,134 (a reduction of roughly 270 times).  Of the 688
(1,090) objects, all 174 known Tycho-2 stars are included, bringing
down the number to search to 514 (916).  135 (163) stars are included
that were flagged as previously catalogued high proper motion objects
(from either Giclas' or Luyten's catalogues).

We can see that we have lost 72 (44) flagged high motion objects,
since there were 207 of them found in the original extraction which
only had a limit on the value of the proper motion.  Of the 72 (44),
23 fall below the $R_2$ brighter than 18.0 magnitude cut-off. Three
more had position errors larger than $350 \, {\rm mas}$ in each
coordinate.  Another seven had proper motion errors larger than $12
{\rm \, mas \, yr^{-1}}$.  An additional 39 (11) were removed by the
requirement that the difference in $R$ magnitudes be less than 0.5
(1.0) magnitude(s).  Magnitude related selection criteria removed the
bulk of the deleted real objects, 62 (34) out of 72 (44), leaving only
10 that were caught by the position and/or motion error criteria.  The
magnitude criteria also removed a very large fraction of the false
objects.  These numbers imply that 65\% (79\%) of the high motion
objects make it through this set of culls, and that as we make the
magnitude match tighter, while we lose more real objects, we also lose
a larger number of not real objects.  Comparing the samples left after
the two magnitude cuts, we are left with only 56\% of the number
candidate stars to check, versus retaining 82\% of the known real high
motion stars in the larger ($|R_1 - R_2| < 1.0$) sub-sample.

\section{Serendipitous New Objects}

As noted above, 774 fields, each $6\arcmin \times 6\arcmin$, were
examined by eye.  71 of the fields contained 82 objects that appeared to
show proper motions by simple examination of the images in sequence.
Because these objects were not the nominal objective of the search,
initially there was no systematic effort to look for other moving
objects in each field.  Once several were noticed, an effort was made to
keep track of them, so in fact the 71 fields were found among a subset
of the 774 fields checked.  Taking a conservative approach, we will
treat 774 as an upper bound on the total area examined.

Each field covers $0.01$ square degrees, meaning we checked $7.74$
square degrees.  Of the possible 82 moving objects, 2 were found to be
not moving upon more careful examination, leaving 80.  Within the set
of 80, there were two Tycho-2 stars (4492-01044-1, and 4133-00625-1),
and 3 stars that were flagged as already known high proper motion
stars (with motions of 0.218, 1.307, $0.263 \, {\rm arcsec \,
yr^{-1}}$).  Thirteen of the stars have proper motions greater than $0.180
\, {\rm arcsec \, yr^{-1}}$, 20 greater than $0.150 \, {\rm arcsec \,
yr^{-1}}$, and 46 greater than $0.100 \, {\rm arcsec \, yr^{-1}}$.
The positions of these objects are plotted in Fig.~\ref{fig-thr},
where the new objects are shown as filled squares, and the previously
identified objects are crosses.  The distribution of proper motions is
shown in Fig.~\ref{fig-srhist}.

Under the simplest assumptions, this implies that there could be at
least 69,000 objects with motions greater than $0.180$ ${\rm arcsec \,
yr^{-1}}$, and on the order of 240,000 objects with detectable motions
above $0.1 \, {\rm arcsec \, yr^{-1}}$.  The number of objects with
motion greater than or equal to $0.180 \, {\rm arcsec \, yr^{-1}}$ is
in line with the number of objects already in the NLTT (just under
$60,000$) which has a nominal lower detection threshold of $0.180 \,
{\rm arcsec \, yr^{-1}}$.

\subsection{Position and Proper Motion Determination}

A quick look at the USNO-B1 catalogue data for these serendipitous
objects led to the realization that about half of the objects had
incomplete or incorrect USNO-B1 entries; detections were mismatched or
missing.  As a result, we decided to re-do the position and proper
motion determinations by hand for all of these objects.

We extracted digitized Schmidt plate material for fields around each
of the 80 objects and re-computed the positions and proper motions of
the moving objects.  Because the scans of the Schmidt plates served by
the USNO Archive server
have not been merged spatially, if a pointing lies in the overlap
region of two plates, the image data from BOTH are available.  For
many of these serendipitous objects, there are more than 5 images of
the field available (from a minimum of 4 plates, to a maximum of 14,
with a mean of 8 plates per field; 65 out of 80 of the objects were in
plate overlap regions).

In each field, a moderate number of nearby stars (within several
arcminutes of the object of interest) with no detectable motion (both by
eye, and per the USNO-B1 catalogue information) were chosen as reference
stars.  We measured their centroids and the centroid of the moving
object and then did a linear plate solution for each set on each plate.
To the measured positions of the moving object, we fit a straight line
for position and proper motion.

Among the caveats to keep in mind, many objects in USNO-B1 have proper
motions of zero, with zero errors.  This indicates that the fit for
position and motion was not very good.  No new work has been done to
correct for the degradation in the astrometric solutions on the
Schmidt plates out near the edges \citep[see][section 4, and Fig.~1
for their discussion of the fixed pattern astrometric errors on the
plates; these rise to the order of arcseconds out near the plate
edges]{metal03}.  Finally, it is important to remember that the proper
motions listed in USNO-B1 are relative proper motions, and the zero
point was set by the least squares solutions for the plates at around
magnitude 18 \citep{metal03}.

\subsection{Results from the serendipitous objects}

Of the 80 objects, 41 had good solutions in USNO-B1.  By virtue of
having re-done the fits for all the objects, we had a reference sample for
the astrometry.  The results of our hand fits were in good agreement
with the numbers given in USNO-B1.  This gave us a certain degree of
confidence in the results for the other 39 for which USNO-B1 does not
have complete or correct data.

Not all of the 39 objects for which we re-did the solutions had bad
data in USNO-B1.  Looking at objects where the USNO-B1 solution was
based on 3 out of 5 possible plates ($N_{FitPts} = 3$), for 6 of them
USNO-B1 has reasonable positions and proper motions, and for another
10 (14), USNO-B1 has a position that is correct to within 2 (4)
arcsec.  For objects where $N_{FitPts} = 4$, all 8 are mis-matched.
For objects where $N_{FitPts} = 2$, all 8 have positions at least 5
arcsec away from the re-computed positions.  In all these cases, when
determining the position and motion by hand, we found the objects on
at least 4 plates.  So, it would be fair to say that USNO-B1 has
reasonable positions and motions for 47 out of the 80 objects (59\%),
and decent positions for 10 more (71\%).

These numbers are not as complete as we might hope, but they are
actually not out of line with the completeness seen by \citet{g03},
though the sample used does not contain many stars with motions slower
than 150 ${\rm mas \, yr^{-1}}$.  Our own check against the complete
revised LHS \citep{bsn02} similarly shows USNO-B1 to be roughly 80\%
complete between 0 and 1 ${\rm arcsec \, yr^{-1}}$, though there are
only a few hundred objects in that catalogue with motions below $500
\, {\rm mas \, yr^{-1}}$ (see the section 6 discussion comparing the
revised LHS with USNO-B1).  In addition, we would be very cautious
about deducing too much about the completeness of USNO-B1 from this
sample, as it suffers from a variety of biases.  First, almost half of
the 80 objects were found in just 3 POSS-I fields, so if a given plate
had some problem, that could affect the results.  Second, when the
objects were found, there was initially no systematic effort to keep
track of them.  Finally, the sample is small.  With those caveats in
mind, it is interesting to note that all of the objects for which we
provide re-done solutions have motions between 0 and 300 ${\rm mas \,
yr^{-1}}$.  J. Munn (private communication) when comparing USNO-B1 to
the SDSS DR1 data sees a dip in completeness from about 95\% to about
65\% at motions of around 80 to 100 ${\rm mas \, yr^{-1}}$, which does
roughly correspond to where we find most of the objects that had
incorrect data in USNO-B1.  A more thorough study of this is
warranted.

From examination of the digitized images, it appears that one of the
primary reasons that almost half of these objects were mis-matched in
the catalogue is that they fall in the overlap zones between fields
(at least 65 of the objects lie in plate overlap regions).  The
detections on adjoining plates apparently were not culled completely
in the duplicate detection removal process.  Since the duplicate
removal depends upon spatial coincidence, and the plate solutions are
at their worst out near the plate edges, this could lead to larger
than expected offsets between images of the same object on different
plates, and hence alternate detections might slip through the
duplicate removal process; the occurrence of multiple entries in
USNO-B1 for the same object and the inability to properly match up
some first and second epoch detections of the same object could also
be explained by this.  This type of error should be most pronounced
among objects with moderate to large proper motion.  This is
potentially important as well because a non-negligible portion of the
sky lies in overlap zones (on the order of 30 to 50\% of the sky).

As noted above, 5 of the objects were flagged in USNO-B1 as previously
known (i.e. in Tycho-2 or one of the high proper motion catalogues).
We checked the rest of the objects against the catalogues and journal
tables made available at CDS\footnote{http://cdsweb.u-strasbg.fr/}.
Another 6 turned up as previously known \citep[including two that comprise
the common proper motion pair LDS~4990;][]{lds}. We have treated
the remaining 69 as previously unknown.  The distribution of total
proper motions is shown in the histograms in Fig.~\ref{fig-srhist}.
Position and motion data for all the objects are given in
Tables~\ref{tbl-serub} and \ref{tbl-sersel}, where
Table~\ref{tbl-serub} has the data for the objects with good solutions
in USNO-B1, and Table~\ref{tbl-sersel} has the information on those
objects which were re-done by hand.

\section{Notes regarding specific objects}

The serendipitous objects were originally numbered 1 through 82
starting with the prefix MUSR  (hence MUSR 01 to MUSR 82).  For those objects
with good positions in USNO-B1, we refer to each by the USNO-B1
designator.  For the objects re-fit by hand, we use the MUSR 
designator.

We constructed a reduced proper motion diagram to aid in the rough
classification of the newly found objects (Fig.~\ref{fig-rpm}).  The
reduced proper motion in the photographic R band is defined as
$$
H_R = R + 5 + 5 \log_{10} (\mu) = M_R + 5 \log_{10}(v_{\rm tan}) - 3.38
$$ where R is apparent magnitude, $M_R$ is the absolute magnitude, and
$v_{\rm tan}$ is the transverse velocity in ${\rm km \, s^{-1}}$.  The
reduced proper motion has the benefit of being insensitive to the
distance to the object, as the distance dependence of $M_R$ and
$v_{\rm tan}$ cancel out.  This has been plotted against $R-K_s$ color
and, as previously shown by \citet{sg02}, and \citet{lrs03a} does a
reasonable job of distinguishing between disk dwarfs, halo subdwarfs
and white dwarfs.  Tentative classifications are given in
Tables~\ref{tbl-ltone} through \ref{tbl-sersel}.  Possible white
dwarfs include objects USNO-B1 1686-0094267 and MUSR 39.  Possible
sub-dwarfs include USNO-B1 1180-0331814, 0484-0243338, 1540-0035963,
1522-0148544, 0867-0255338, 1663-0069093 and MUSR 40, MUSR 54, MUSR 65
and MUSR 82.

{\it LHS~237a (0560-0118956):\/} This object was originally thought to
be new.  Upon closer inspection (H. Harris, private communication),
it was found to be LHS~237a (or VBs3).  The LHS position given for
this is off by $8\arcmin$ in declination.  The RA and the proper
motion both match.  The finder given in \citet{vb61} matches the
images of this object.  In \citet{bsn02}, this object is listed as not
found.  Correct positions and motions are given in
Table~\ref{tbl-known} (see Fig~\ref{fig-hm0560}).

\subsection{Objects with relatively large proper motion in galactic latitude}

In a modest effort to point out stars that might be halo stars,
we have singled out objects that meet the following criteria:
$\mu > 0.75 \arcsec \, {\rm yr^{-1}}$ and $\mu_b > 2 \mu_l$.

{\it 0258-0023144:\/} At $(l, b) = 278{\fdg}5505, -44{\fdg}0139$ moving along
$(\mu_l, \mu_b) = 391.8, 999.8 \, {\rm mas \, yr^{-1}}$ (Fig.~\ref{fig-mub}(a) and
Table~\ref{tbl-known}).

{\it 0867-0249298:\/} At $(l, b) = 269{\fdg}6031, 54{\fdg}7070$ moving along
$(\mu_l, \mu_b) = -357.1, -1037.0 \, {\rm mas \, yr^{-1}}$ (Fig.~\ref{fig-hm0867} and
Table~\ref{tbl-new}).

{\it 1657-0005791 (MUSR 06):} At $(l, b) = 121{\fdg}5699, 12{\fdg}8853$ moving along
$(\mu_l, \mu_b) = 39.8, 159.2 \, {\rm mas \, yr^{-1}}$ (Fig.~\ref{fig-mub}(b) and
Table~\ref{tbl-serub}).

{\it 1540-0035963 (MUSR 12):} At $(l, b) = 124{\fdg}9866, 1{\fdg}2375$ moving along
$(\mu_l, \mu_b) = 2.6, 128.5 \, {\rm mas \, yr^{-1}}$ (Fig.~\ref{fig-mub}(c) and
Table~\ref{tbl-serub}); possible sub-dwarf.

{\it 1570-0182321 (MUSR 43):} At $(l, b) = 98{\fdg}7379, 37{\fdg}8952$ moving along
$(\mu_l, \mu_b) = -6.5, 100.9 \, {\rm mas \, yr^{-1}}$ (Fig.~\ref{fig-mub}(d) and
Table~\ref{tbl-serub}).

{\it 1544-0281760 (MUSR 60):} At $(l, b) = 109{\fdg}9080, 4{\fdg}8349$ moving along
$(\mu_l, \mu_b) = -21.4, -83.7 \, {\rm mas \, yr^{-1}}$ (Fig.~\ref{fig-mub}(e) and
Table~\ref{tbl-serub}).

{\it 1558-0247969 (MUSR 69):} At $(l,b) = 112{\fdg}3325, 5{\fdg}2338$ moving along
$(\mu_l, \mu_b) = -50.2, 104.9 \, {\rm mas \, yr^{-1}}$ (Fig.~\ref{fig-mub}(f) and
Table~\ref{tbl-serub}).

\subsection{Objects with companions}

In the process of putting together the tables and images of
serendipitous objects, we noticed several pairs with very similar
motions.  They are listed here.

{\it MUSR 40:} There is a possible faint companion to the east of this
object that is visible on the POSS-I 103aO (blue) plate for field 68,
and on the POSS-II IV-N (near-IR) plate for field 67 (see
Fig.\ref{fig-sr40}(a) and (c)).  The faint companion is at the same
position angle and distance with respect to MUSR 40 on both plates,
though they were taken over 40 years apart.  The POSS-I 103aO plate
for field 69 also shows something peculiar near MUSR 40
(Fig~\ref{fig-sr40}(b)).  The object is not seen on the other POSS-I
and POSS-II plates that cover this object.  The corresponding 103aE
images for both POSS-I images are shown as Fig~\ref{fig-sr40}(d) and
(e), and an additional POSS-II IV-N image from field 68 is shown in
Fig~\ref{fig-sr40}(f) (see Table~\ref{tbl-sersel}).  MUSR 40 is a possible
sub-dwarf.

{\it LDS~4990 (MUSR 56) and 1543-0282460 (MUSR 57):} The two
components of LDS~4990 \citep{lds} are shown in
Fig.~\ref{fig-lds4990}, with data in Tables~\ref{tbl-serub} and
\ref{tbl-sersel}. In Fig.~\ref{fig-lds4990}, MUSR 56 is marked with a
circle on both images, MUSR 57 is marked with a square, and MUSR 58
(see next entry) is marked with an ellipse.  We note in passing that
MUSR 56 is coincident to within 3 arcseconds with 1RXS
J224000.2+642310 (marked with a white X on Fig.~\ref{fig-lds4990}(b))
\citep{rxs}.

{\it 1543-0282475 (MUSR 58):} Very near to LDS~4990.  This object's
motion is in the same direction as that of the components of LDS~4990,
but the magnitude of the motion in markedly smaller (see
Fig~\ref{fig-lds4990} and Table~\ref{tbl-serub}).

{\it MUSR 61 and MUSR 62:} A probable co-moving pair, with a separation of
roughly 58 arcseconds (see Fig.~\ref{fig-sr61_62} and
Table~\ref{tbl-sersel}).  MUSR 61 is marked with a circle, and MUSR 62
is marked with a square.

{\it MUSR 76 and MUSR 77:} A probable co-moving pair, with a separation of
almost 6.6 arcminutes (see Fig~\ref{fig-sr76_77} and
Table~\ref{tbl-sersel}).  MUSR 76 is marked with a circle, and MUSR 77
is marked with a square.
They have very similar magnitudes in both the
optical and near IR, and their motions are quite similar as well.

{\it MUSR 81:} Its motion is very similar to that of MUSR 76 and MUSR 77
(Table~\ref{tbl-sersel}), but is yet further separated from the
previous two objects (about $57\arcmin$ away), and is somewhat
brighter than either one.

\section{Comparison with the High Motion part of rLHS and LSR}

As part of our effort to understand how well the motion finder has
done, we looked at the entries in the USNO-B1 catalogue for all the
objects with motions between 1.0 and 5.0 ${\rm arcsec \, yr^{-1}}$ in
the revised LHS catalogue \citep[hereafter rLHS]{bsn02}, and the 18
new objects that meet this criterion found by \citet[][LSR]{lsr02}.

For each of the 18 objects in LSR with motions between 1.0 and 2.0
${\rm arcsec \, yr^{-1}}$, we extracted the appropriate portion of
USNO-B1, and images from the Schmidt photographic surveys that cover
that object.  Of the 18 objects, 7 were matched in USNO-B1 (the seven
found in our search, and given in Table~\ref{tbl-known}).  Of the
other 11, 3 were in fields confused enough that we had only modest
expectations that we would have found them.  One object was on a
diffraction spike of a brighter object, and likely would not have been
found by USNO-B1 because it would have been in a removed region.  Seven
of the 11 objects we should have found.  In several of those cases, it
looked like USNO-B1 matched up the wrong set of objects among the
various survey epochs.  It appears this happens because there are
other objects near or along the line of motion that cause the code
that predicts the motion to get confused.  The LSR image difference
method is complementary to the ``comparison of detection lists''
method used for USNO-B1.  We would expect that LSR should be more
sensitive to objects in highly confused areas (such as the Galactic
plane).

The rLHS has 593 objects with motions between 1.0 and 5.0 ${\rm arcsec
\, yr^{-1}}$.  We found that 171 of these objects are flagged in
USNO-B1 as being Tycho-2 stars (these were added to USNO-B1 directly
from Tycho2) and so don't tell us how well \citet{metal03} did in the
construction of USNO-B1.  For the remaining 422 rLHS stars, we
compared the rLHS proper motions with those given in USNO-B1.  197 had
proper motions that matched within 0.20 ${\rm arcsec \, yr^{-1}}$ and
20 degrees position angle (though most are much closer).  Of these,
174 are flagged in USNO-B1 as being known high proper motion stars.
23 more are matched to other USNO-B1 entries, though they are not
flagged as known motion stars.

There were 225 objects that did not have proper motions matched within
the above limits.  For the 158 of them with LHS catalogue number less
than or equal to 552, we searched a $6\arcmin$ square box around their
position.  For the other 67, we searched a $3\arcmin$ square box.
Fig.~\ref{fig-ublhs-rat}(right upper) and (right middle) panels show the
distribution of proper motions of the matched and un-matched sets of
rLHS objects respectively.  The lower panel shows the percent of rLHS
objects that were matched as a function of proper motion.  
For objects with motions between 1 and 2 ${\rm arcsec \, yr^{-1}}$,
the mix is pretty even.  Above 2 ${\rm arcsec \, yr^{-1}}$, more
objects are not matched (though we are getting into the realm of small
number statistics).

Similar data are shown in the left hand panel of
Fig.~\ref{fig-ublhs-rat} for objects in the rLHS with motions between
0 and 1 ${\rm arcsec \, yr^{-1}}$.  For objects with motions below
about 400 ${\rm mas \, yr^{-1}}$, the completeness appears to dip a
bit, but the sample size per bin is much smaller than for those bins
with proper motions above 500 ${\rm mas \, yr^{-1}}$ (which is not
surprising, given that the catalogue is only supposed to contain
objects with motions larger than 500 ${\rm mas \, yr^{-1}}$).

The USNO-B1 catalogue has decent matches for the position and motion
of 47\% of the rLHS objects with motions between 1.0 and 5.0 ${\rm
arcsec \, yr^{-1}}$.  For these objects, the median distance between
the rLHS and USNO-B1 positions is about 1.9 arcseconds
(Fig.~\ref{fig-ub_lhs}).  This displacement is consistent with what
the typical uncertainty in position in the rLHS which is about 2
arcseconds \citep{bsn02}.  Fig.~\ref{fig-ub_lhs}(b) and (c) shows the
displacement between matched rLHS and USNO-B1 objects in total proper
motion and position angle.  The median difference in proper motion was
0.03 ${\rm arcsec \, yr^{-1}}$, and the median position angle
difference was 1.4 degrees.

\citet{g03} has recently undertaken a more extensive comparison of
USNO-B1 with their revised version of NLTT \citep{gs03,sg03}.  As noted
above, they found USNO-B1 to be roughly 30\% incomplete when $\mu = 1
\, {\rm arcsec \, yr^{-1}}$; that the incompleteness should get worse
as the motion increases above this is a natural assumption.

In preliminary testing of the moving object finding algorithm used in
the construction of USNO-B1, we found that below $1\,{\rm arcsec \,
yr^{-1}}$, the object finding algorithm did substantially better at
finding real motions than it did for the faster moving objects.  Since
a much greater percentage of the moving objects move more slowly than 1
arcsecond per year, even though we appear to have missed many with
large motions, this is consistent with the work of \citet{g03}.

\section{Discussion}

Out of 187,134 objects in USNO-B1 that had listed motions between 1.0
and 5.0 ${\rm arcsec \, yr^{-1}}$, there are 207 objects in USNO-B1
with the flag bit set indicating that they match a high proper motion
catalogue star (no cuts have been applied to these yet).  Of those,
184 have a second epoch red magnitude less than or equal to 18, and
174 are matched from the LHS.  There are another 23 unflagged objects
that match LHS objects, 19 that were recently found in other searches
and 2 new ones, for a total of 251 objects.  There are another 174
Tycho-2 stars with motions in this range that were added in.
Excluding the added Tycho-2 stars, 0.1\% of the objects in USNO-B1 in
this range are real.  It seems fair to say that it is possible to find
new high motion objects in the USNO-B1 catalogue, even with the large
contamination fraction, though it is not easy.  Given that we found
just under half of the previously known high motion objects, and then
also found another two new ones, this would imply that there should be
at least another few waiting to be found.

In addition, we found another 80 objects in the fields we searched for
high motion objects.  For almost half, we had to match up the
detections by hand and compute positions and proper motions.  Out of
the combined high motion and serendipitous sample, seven objects have
motions with relatively large $\mu_b$, and there are four pairs that
appear to be common proper motion pairs, and maybe even one common
motion triple.  In the end, we found 2 new stars with proper motions
larger than 1 ${\rm arcsec \, yr^{-1}}$, and 36 with proper motions
between 0.1 and 1 ${\rm arcsec \, yr^{-1}}$.  We also recovered one
previously known, but recently missed star (LHS 237a) with a motion of
$1.67 \, {\rm arcsec \, yr^{-1}}$.

Applying several simple cuts to the catalogue reduces the number of
false objects dramatically.  (1) Require each object to have a
positional error in each coordinate less than 0.999 arcsec (or
smaller, e.g. less than 0.350 arcsec). (2) Require each object to have
a small proper motion error (less than 12 ${\rm mas \, yr^{-1}}$). (3)
Limit objects to those where the difference between the $R_1$ and
$R_2$ magnitudes is less than 1.0 or 0.5 magnitudes.  These cuts alone
can reduce the contamination in the returned data by several orders of
magnitude. (4) Require each object to be detected on 4 or 5 out of 5
surveys.

By placing a limit on the position and motion errors, we are putting a
tight constraint upon the acceptable matches, since we are
imposing a linearity requirement in addition to the proximity
criterion.  Hence, the much greater reduction in the number of
spurious objects.  Fig.~\ref{fig-diff} shows an example of this.  The
left panel shows a POSS-I image of a field near a bright star.  In the
center panel, all of the objects in USNO-B1 that lie in this field are
plotted (they number 704).  If we require $|R_1 - R_2| \le 1$, the
total $\sigma_{\rm position} \le 500 \, {\rm mas}$ and the total
$\sigma_{\mu} \le 100 \, {\rm mas \, yr^{-1}}$, then we are left with the 163
objects overplotted in the right panel.  Almost 75\% of the objects
have been rejected by this cut.  As can be seen, most of the artifact
objects caused by the diffraction spikes and the halo around the star
are gone.  A few real objects have been deleted as well.

Requiring objects to be detected on at least 4 surveys did not appear
to contribute much to reducing the contamination in the high motion
sample.  I would attribute this to several factors.  First, the
diffraction spikes on plates taken at the same pointing tend to line
up well (hence the fairly large number of objects discarded as being
due to diffraction spikes), and so provide large pool of objects close
together at both epochs.  These then often project onto or very
near to other diffraction spike detections, thus making up spurious,
though complete, objects with potentially large motions.
Second, extended objects, much like diffraction spikes often give
rise to multiple detections all in close proximity to each other.
These again provide fertile territory for mis-matching.

The high motion problem is particularly taxing for the object
matching, since there are often very many possible pairings of
objects.  With the larger motions, it becomes more likely that
something will fall within the large projected error ellipse, and
hence make up an object with at least 4 detections.

\begin{figure}
\plotone{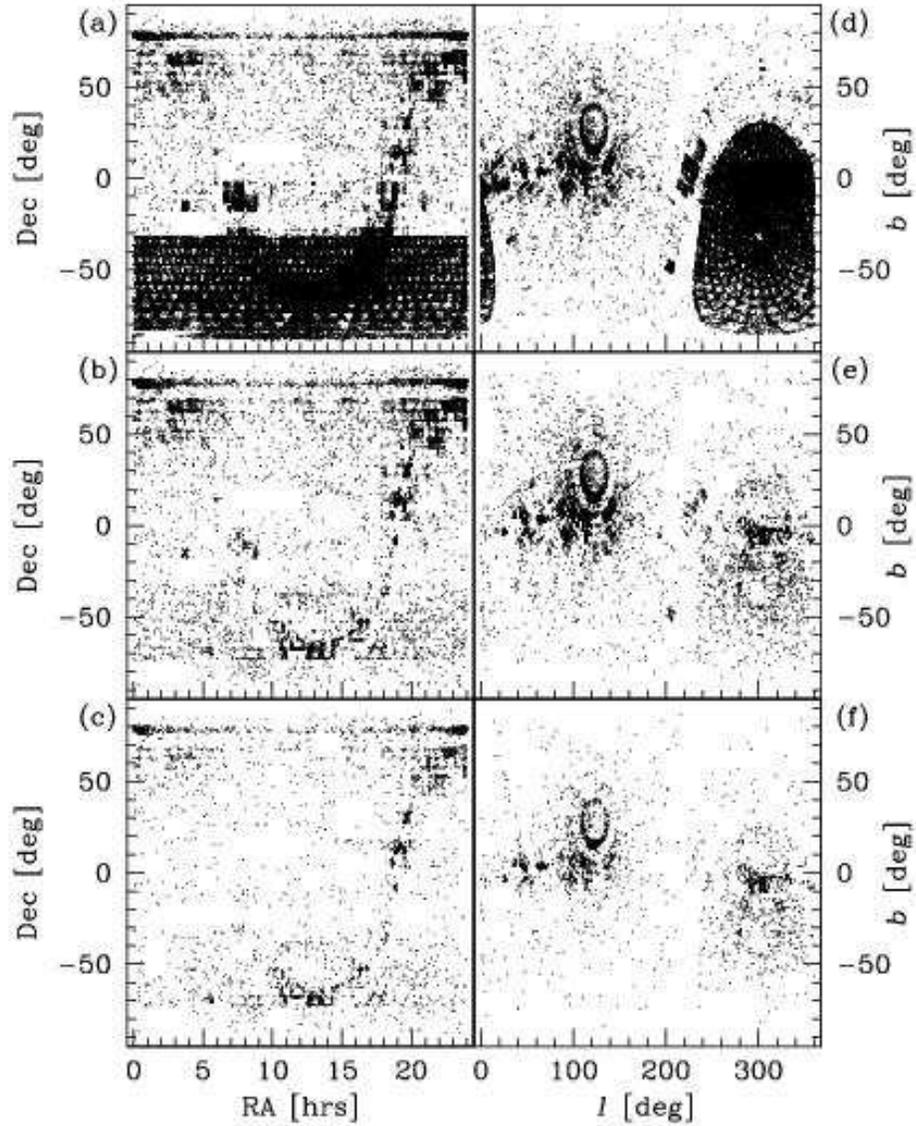}
\caption{Positions of candidate objects shown in equatorial (panels a,
  b, c) and galactic (panels d, e, f) coordinates.  In (a) and (c) are
  all the objects in USNO-B1 with $1 \le \mu \le 5 \, {\rm arcsec \,
  yr^{-1}}$.  The points in (b) and (e) are those remaining after
  basic sanity checks have been applied ($N_{FitsPts} \ge 4$,
  $\sigma_\alpha < 999 \,{\rm mas}$ and $\sigma_\delta < 999 \, {\rm
  mas}$).  Points in (c) and (f) are those remaining after a
  subsequent cut on the second epoch red magnitude ($0 \le {\rm R_2}
  \le 18.0$).
 \label{fig-one}}
\end{figure}

\begin{figure}
\epsscale{0.3}
\plotone{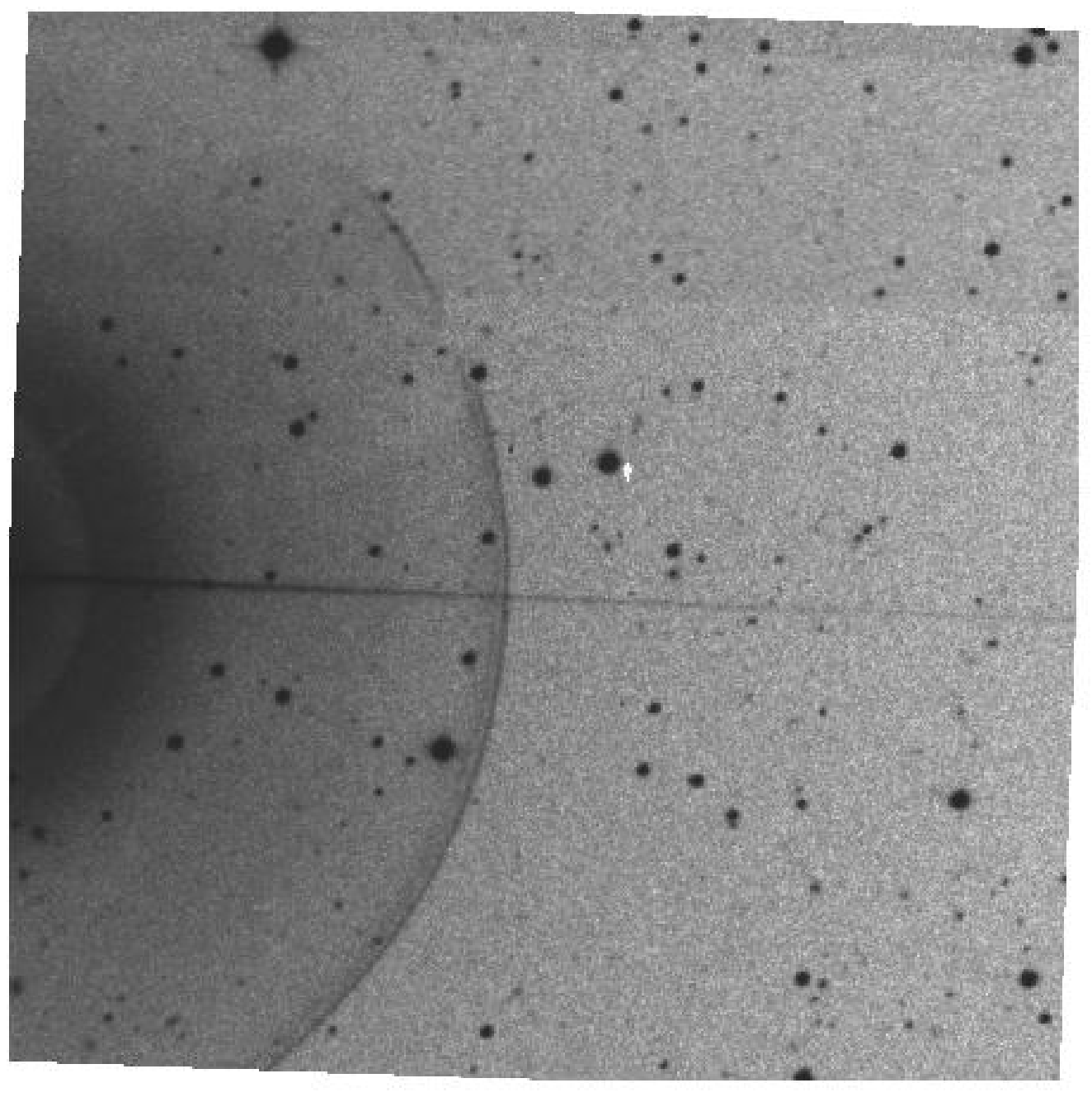}
\plotone{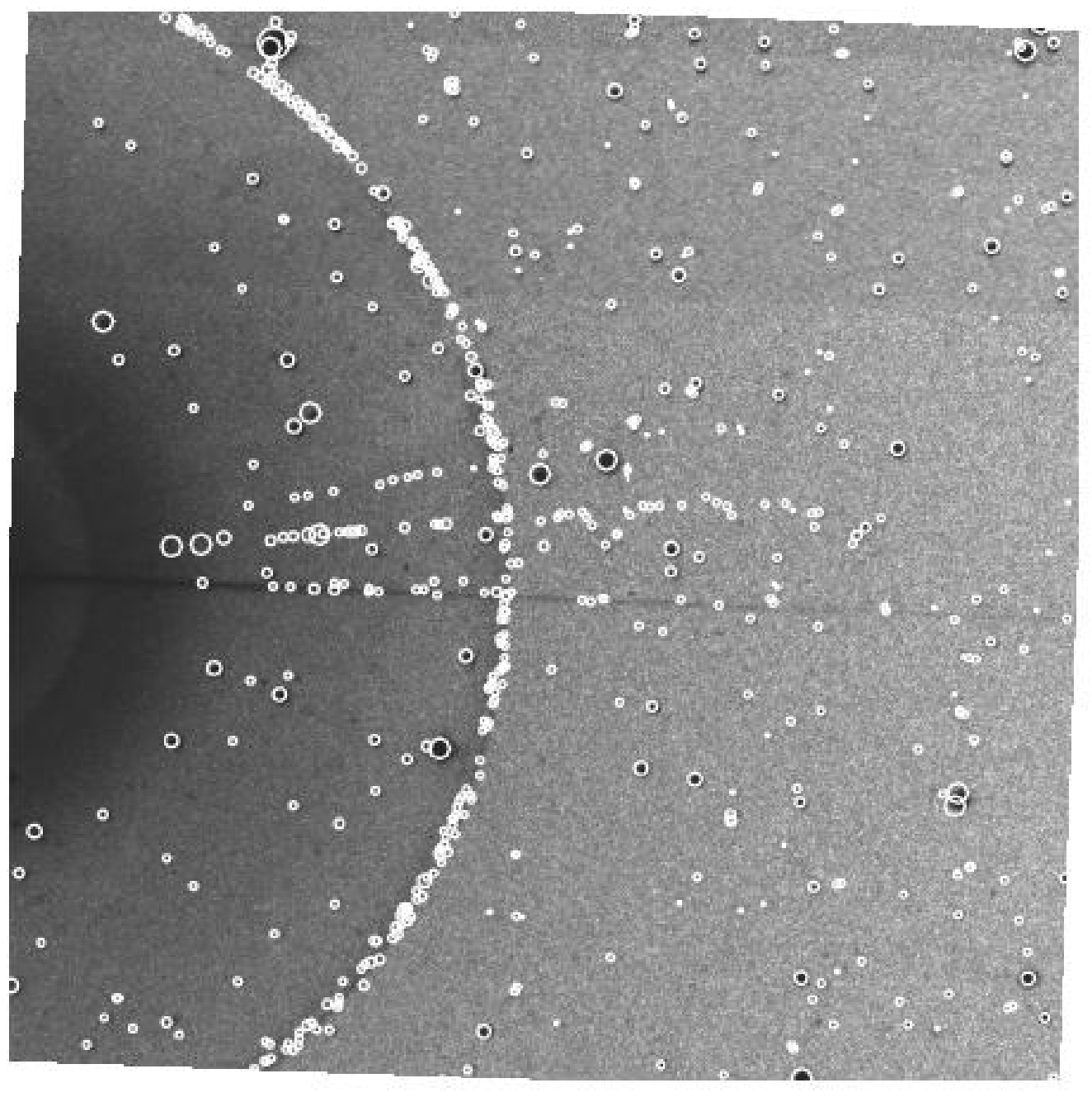}
\plotone{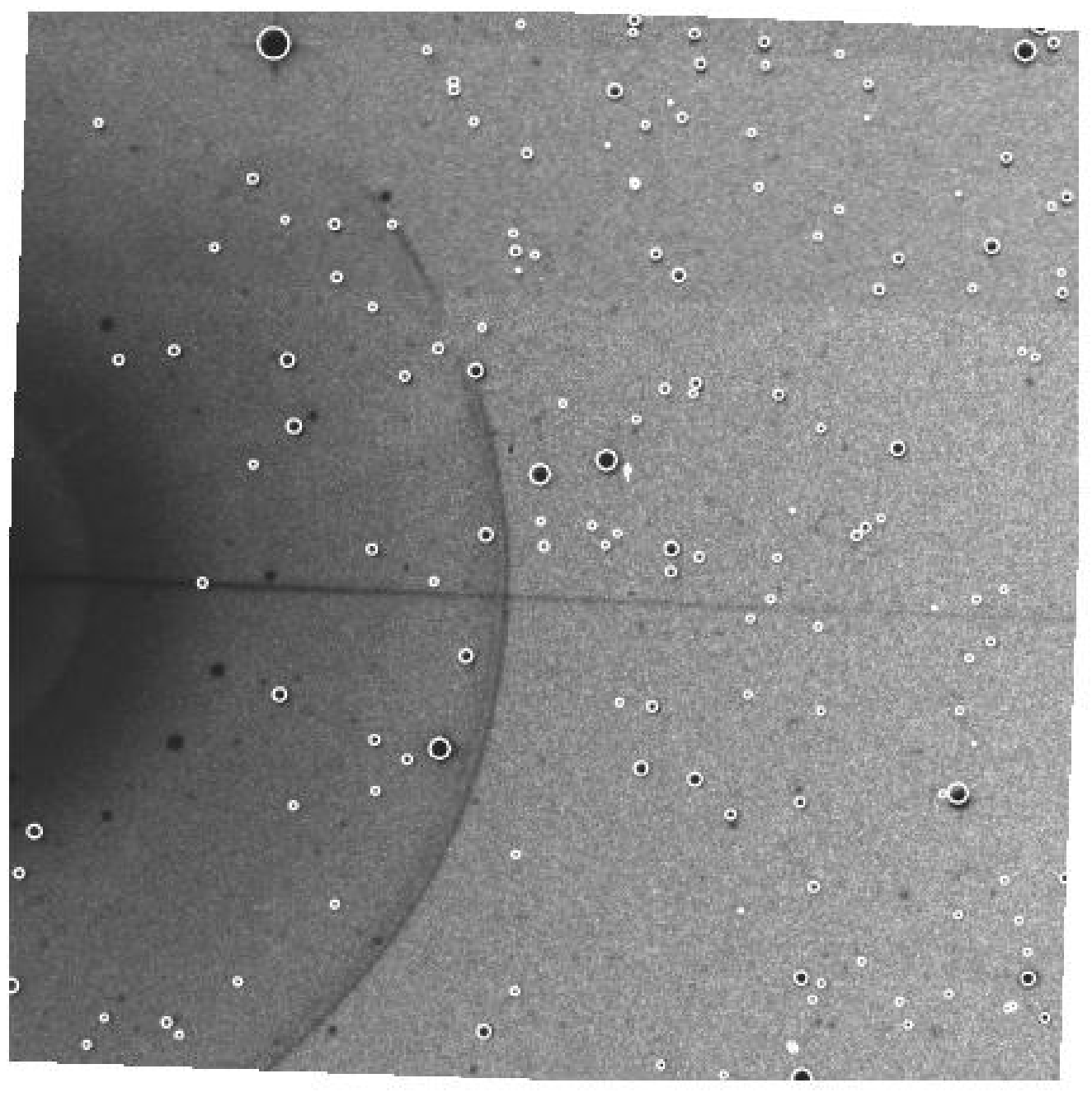}
\caption{{\it left:\/} The POSS-II IIIaF image of a field near a
  bright star. {\it center:\/} The same image, with all of the 704
  USNO-B1 objects that lie in the field overplotted.  {\it right:\/}
  Only those 163 USNO-B1 objects that satisfy the criteria $|R_1 -
  R_2| \le 1 \, {\rm mag}$, $\sigma_{\rm pos} \le 500 \, {\rm mas}$,
  and $\sigma_\mu \le 100 \, {\rm mas \, yr^{-1}}$ are overplotted.
\label{fig-diff}}
\epsscale{1.0}
\end{figure}

\begin{figure}
\plotone{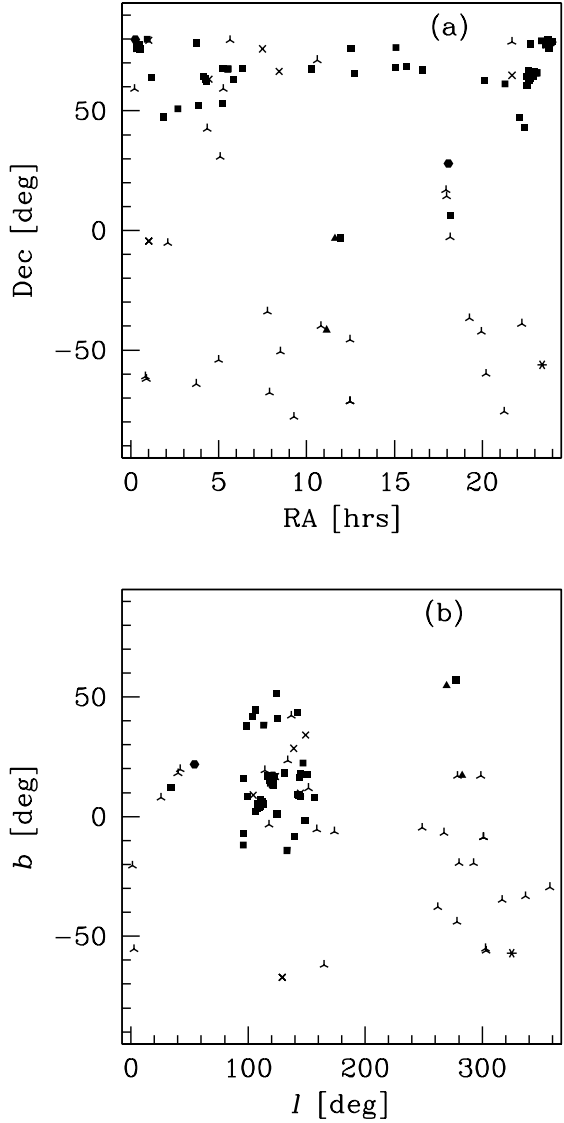}
\caption{The high motion objects that made it all the way through the
  winnowing process.  Filled triangles and 3 pointed stars represent
  objects with proper motion larger than 1 arcsecond that are
  respectively new, and known but not flagged in USNO-B1.  Filled
  hexagons and 6 pointed stars are are objects listed in USNO-B1 with
  large motions that actually have motions less than 1 arcsecond, and
  are new and previously known respectively.  Filled squares and
  crosses represent respectively new and known serendipitous objects.
  Panels (a) and (b) show the objects in equatorial and galactic
  coordinates respectively.
 \label{fig-thr}}
\end{figure}

\begin{figure}
\plotone{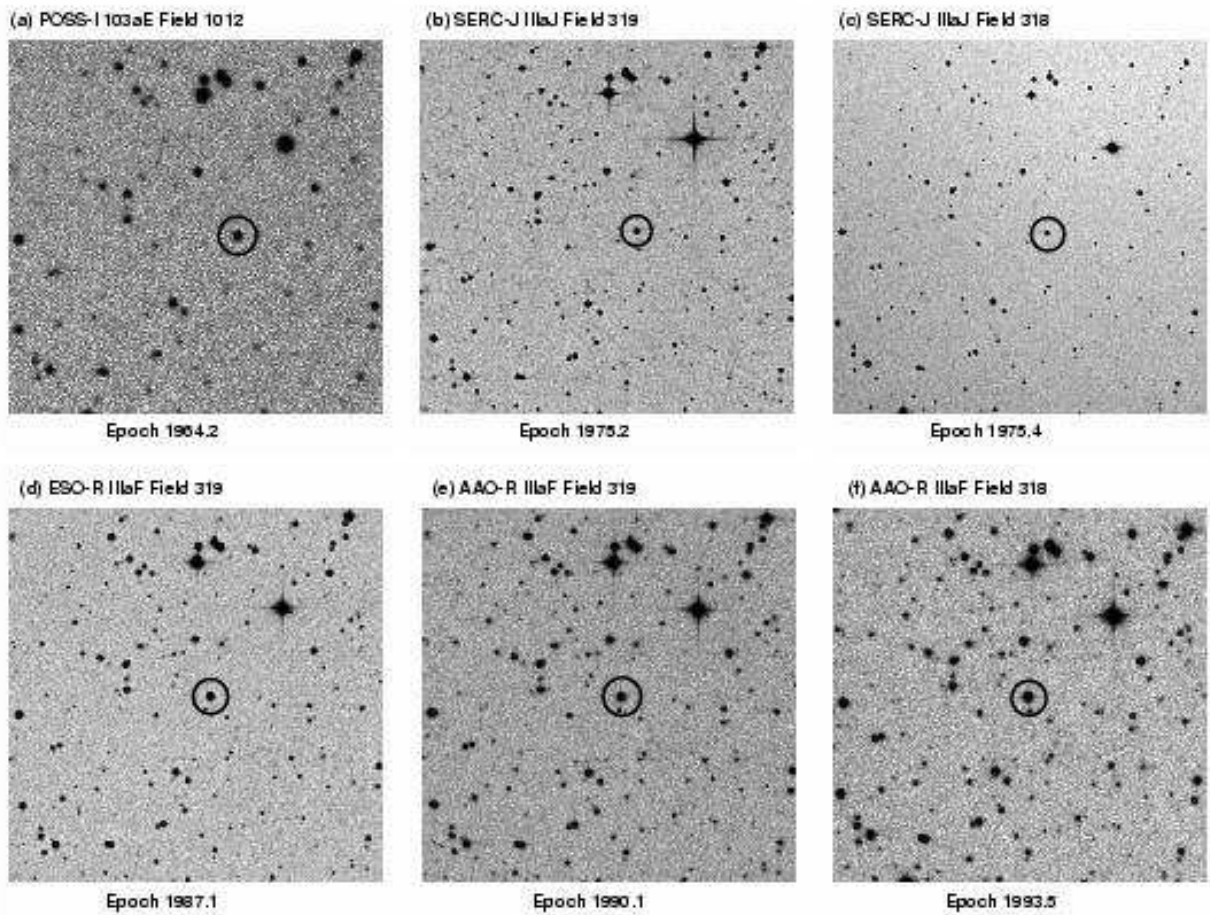}
\caption{Finder charts showing the motion of object USNO-B1
0484-0243338 ($\alpha_{2000} = 11^{\rm h}07^{\rm m}55{\fs}9$,
$\delta_{2000} = -41{\degr}35{\arcmin}53\arcsec$) on the 6
available Schmidt plates.  North is up, East is to the right and all
images are $6{\arcmin}{\times}6{\arcmin}$ in size.\label{fig-hm0484}}
\end{figure}

\begin{figure}
\plotone{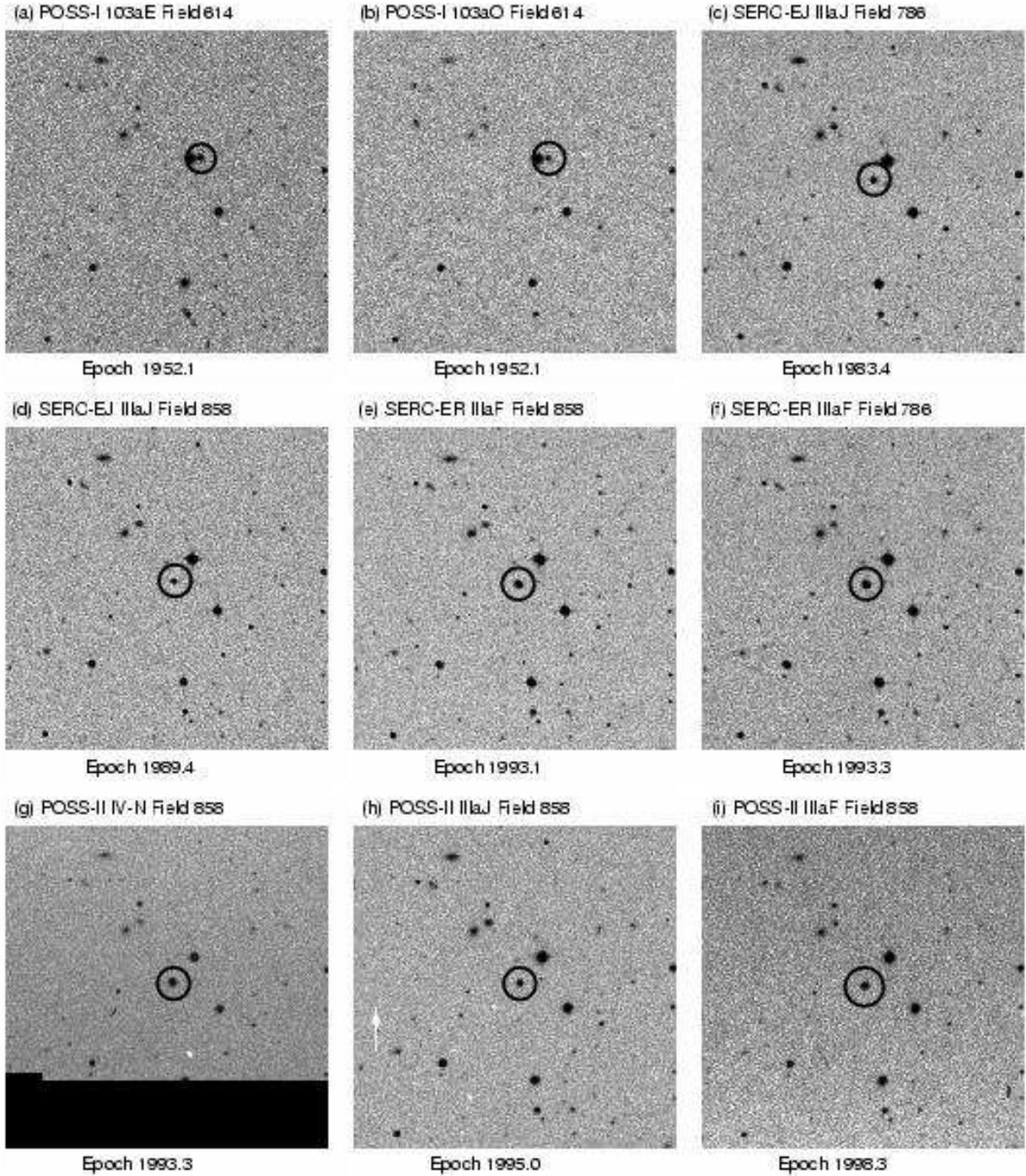}
\caption{Finder charts showing the motion of object USNO-B1
0867-0249298 ($\alpha_{2000} = 11^{\rm h}37^{\rm m}16{\fs}6$,
$\delta_{2000} = -03{\degr}17{\arcmin}37{\arcsec}$) on the 9
available Schmidt plates.  North is up, East is to the right and all
images are $6{\arcmin}{\times}6{\arcmin}$ in size.\label{fig-hm0867}}
\end{figure}

\begin{figure}
\plotone{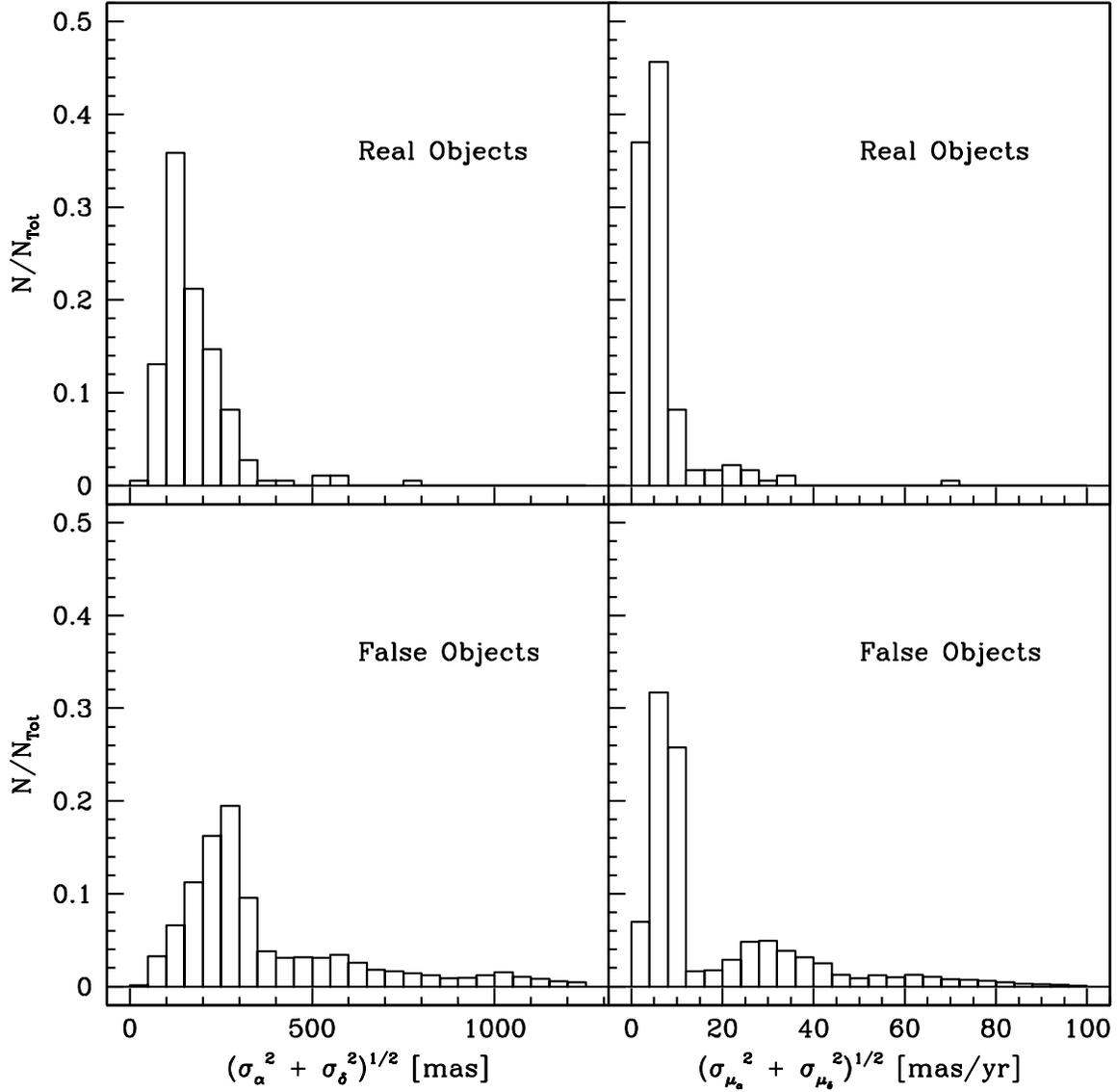}
\caption{Histograms of the errors for the positions ({\it left
  column\/}) and motions ({\it right column}) for the real ({\it top
  row\/}) and false ({\it bottom row\/}) high motion
  objects. This is for the sub-sample of 3,348 possible objects where the
  catalogue based finders were reviewed by eye.\label{fig-hisigs}} 
\end{figure}

\begin{figure}
\plotone{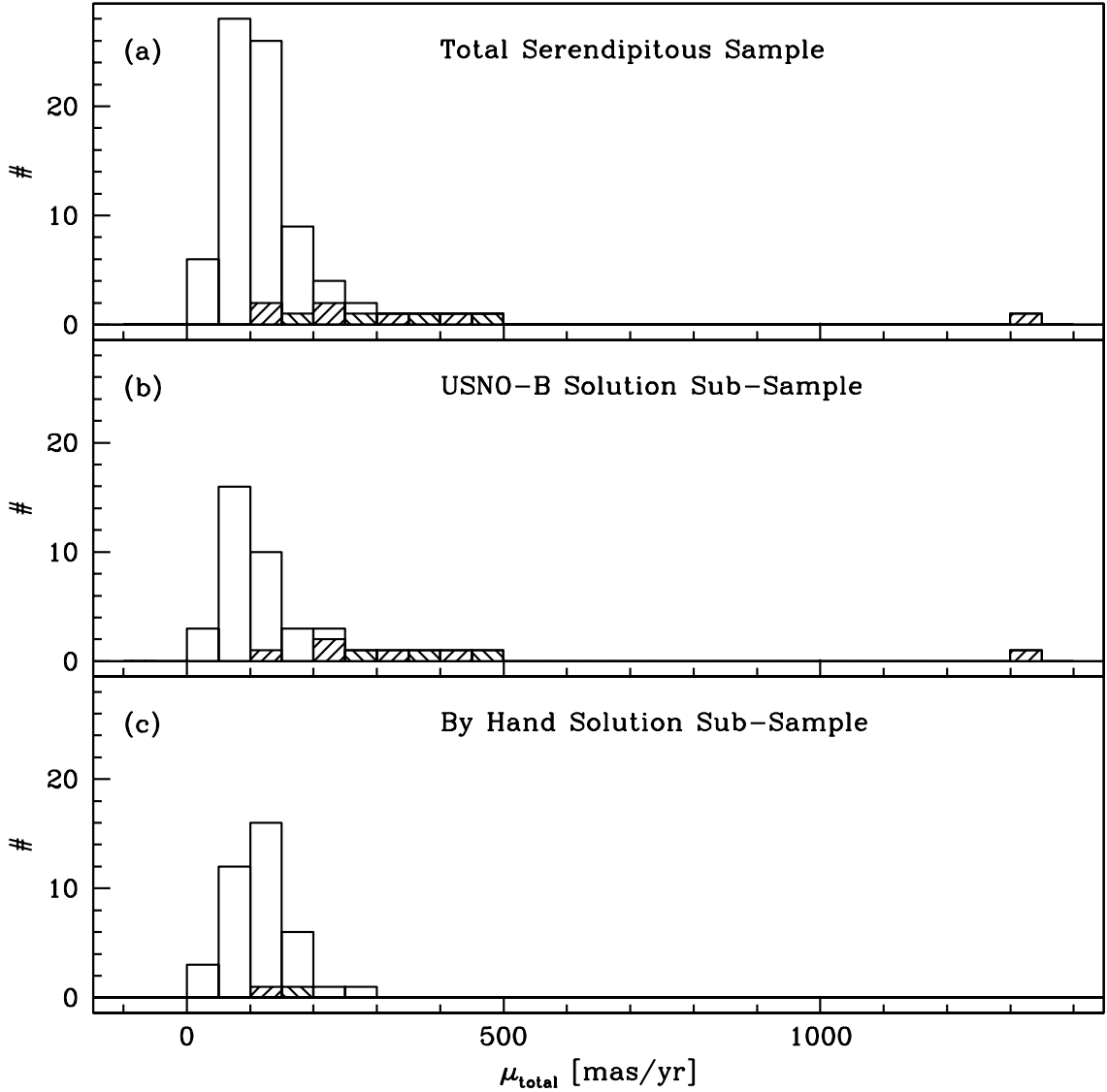}
\caption{Histograms of the total proper motions of the serendipitous
  objects.  The shaded histograms are of the known objects, while the
  outline histograms show the combined new and known.  Panel (a) is the 
  total set of serendipitous objects. (b) is the distribution of objects
  with good solutions in USNO-B1. (c) is the set of objects with
  solutions done by hand.
\label{fig-srhist}} 
\end{figure}

\begin{figure}
\plotone{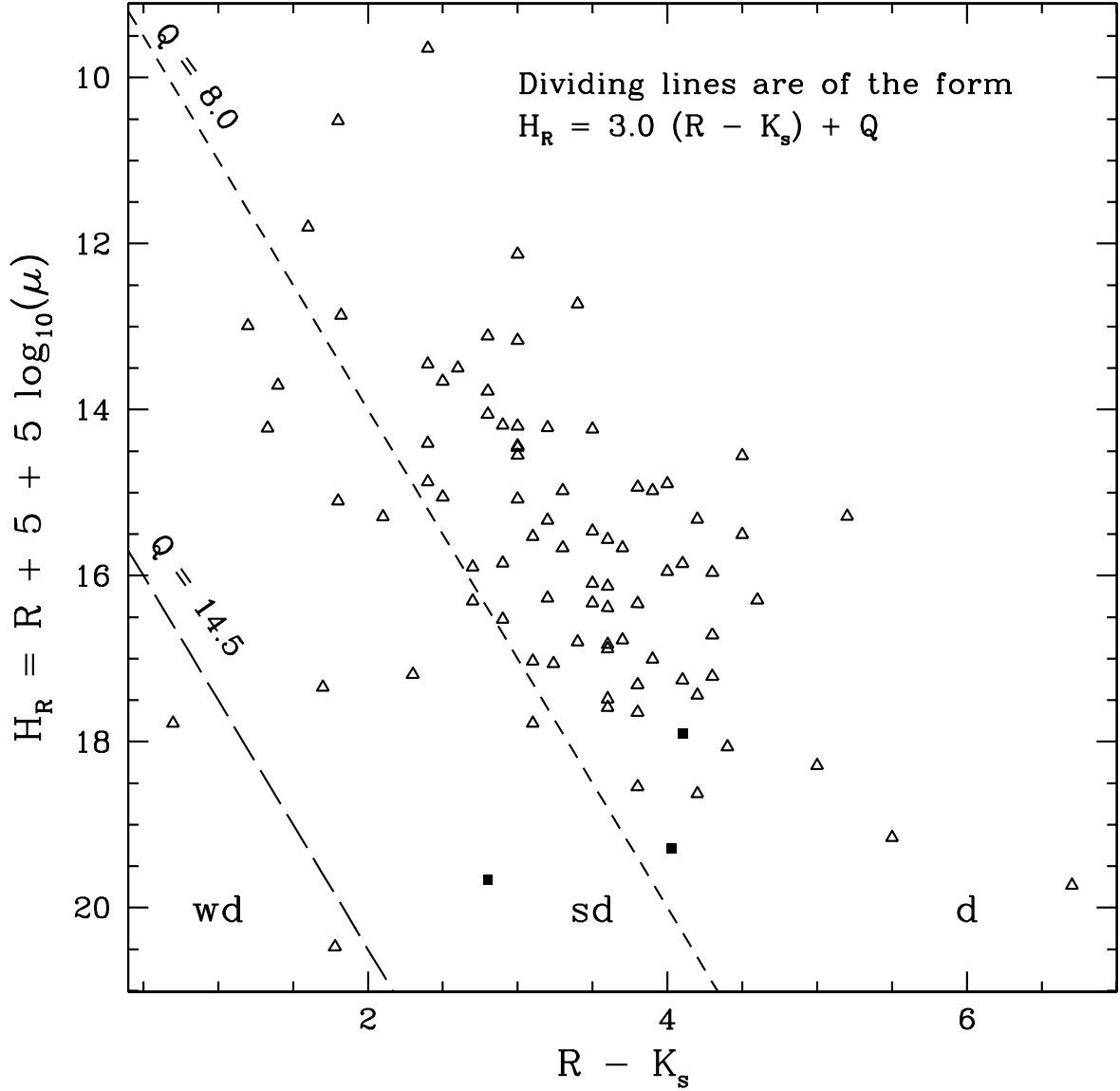}
\caption{Reduced proper motion diagram for the objects listed in
  Tables~\ref{tbl-ltone} through \ref{tbl-sersel}.  The reduced proper
  motion is $H_R = R + 5 + 5 \log(\mu[{\rm arcsec \, yr^{-1}}])$.  The
  lines dividing the space into regions occupied by dwarfs (d),
  sub-dwarfs (sd) and white dwarfs (wd) are based on the work of
  \citet{lrs03a,lsr03b}, and \citet{l05}, and give a rough guide to
  likely stellar type.  Objects with motions larger that $1 \, {\rm
  arcsec \, yr^{-1}}$ are marked with filled squares.
\label{fig-rpm}}
\end{figure}

\begin{figure}
\plotone{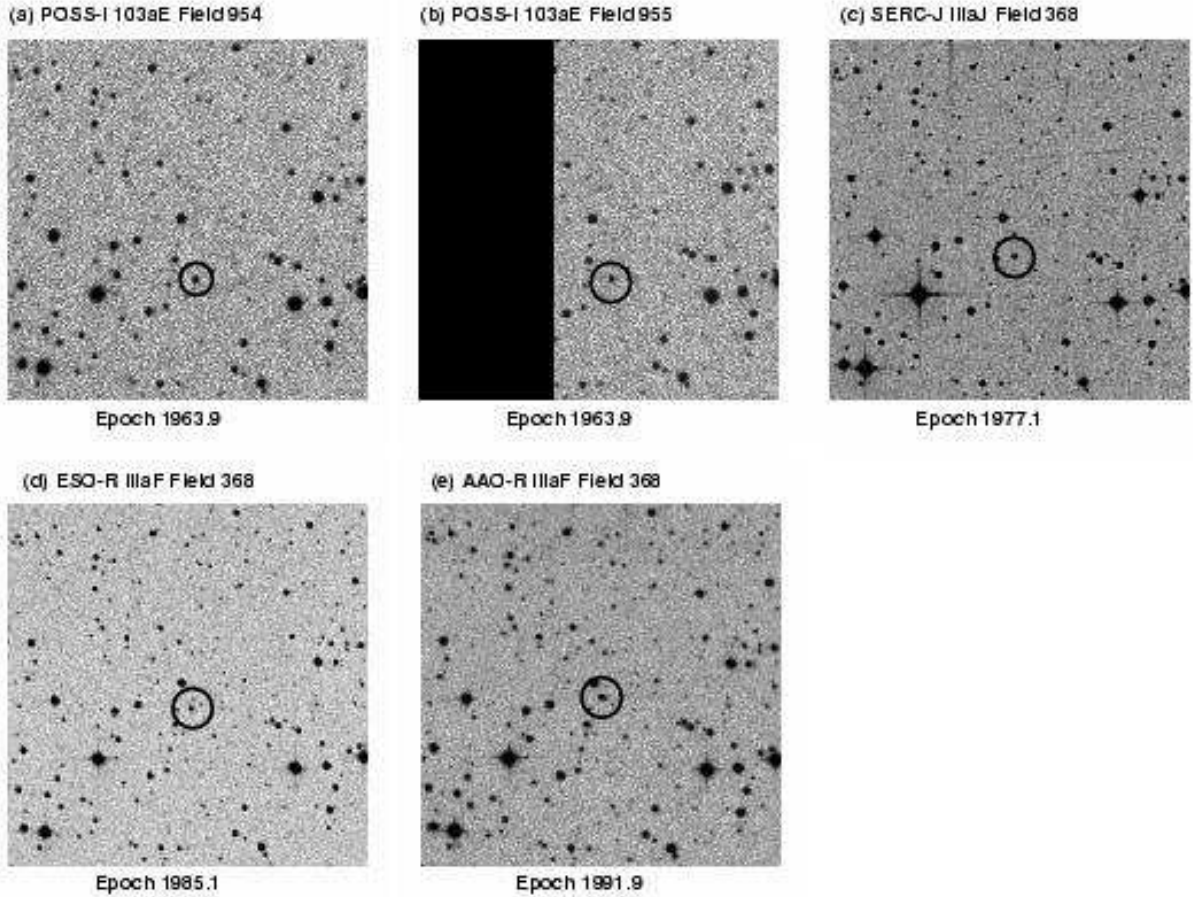}
\caption{Finder charts showing the motion of object LHS~237a
  (USNO-B1~0560-0118956) 
  ($\alpha_{2000} = 07^{\rm h}45^{\rm m}38{\fs}5$, 
  $\delta_{2000} = -33{\degr}55{\arcmin}52{\arcsec}$) on  
  the 5 available Schmidt plates.  North is up, East is to the right
  and all images are $6{\arcmin}{\times}6{\arcmin}$ in
  size.
\label{fig-hm0560}}
\end{figure}

\begin{figure}
\plotone{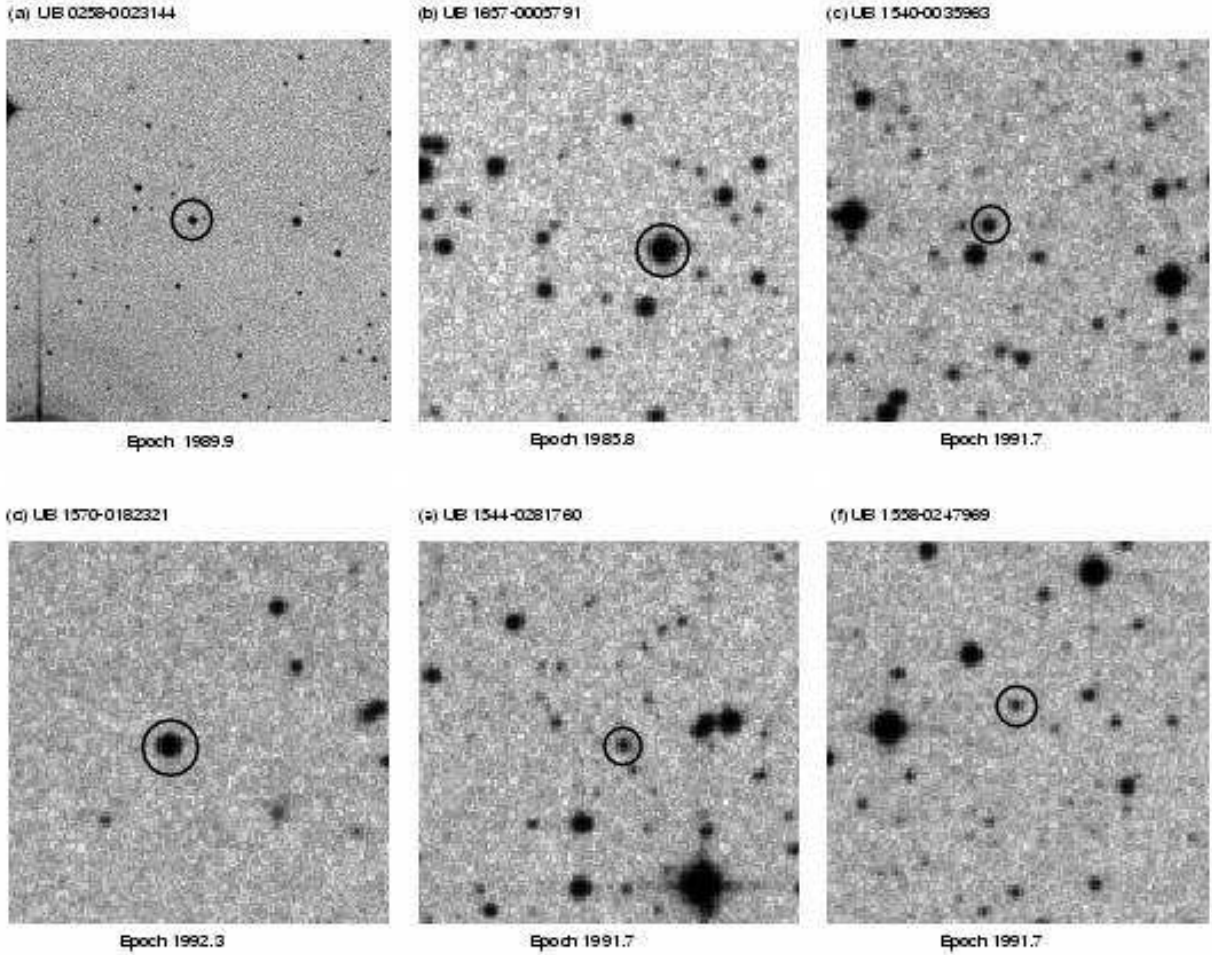}
\caption{Finder charts showing the 6 objects with large relative
  $\mu_b$.  Image (a) is $6{\arcmin}$ square, and images (b) - (f) are
  $2{\arcmin}$ square.  North is up, East is to the right.  Image (a)
  is from an AAO-R IIIaF plate and images (b)--(f) are from POSS-II
  IIIaF plates. (a) USNO-B1 0258-0023144, (b) USNO-B 1657-0005791, (c)
  USNO-B 1540-0035963 (d) USNO-B 1570-0182321, (e) USNO-B 1544-0281760
  (f) USNO-B 1558-0247969.
\label{fig-mub}}
\end{figure}

\begin{figure}
\plotone{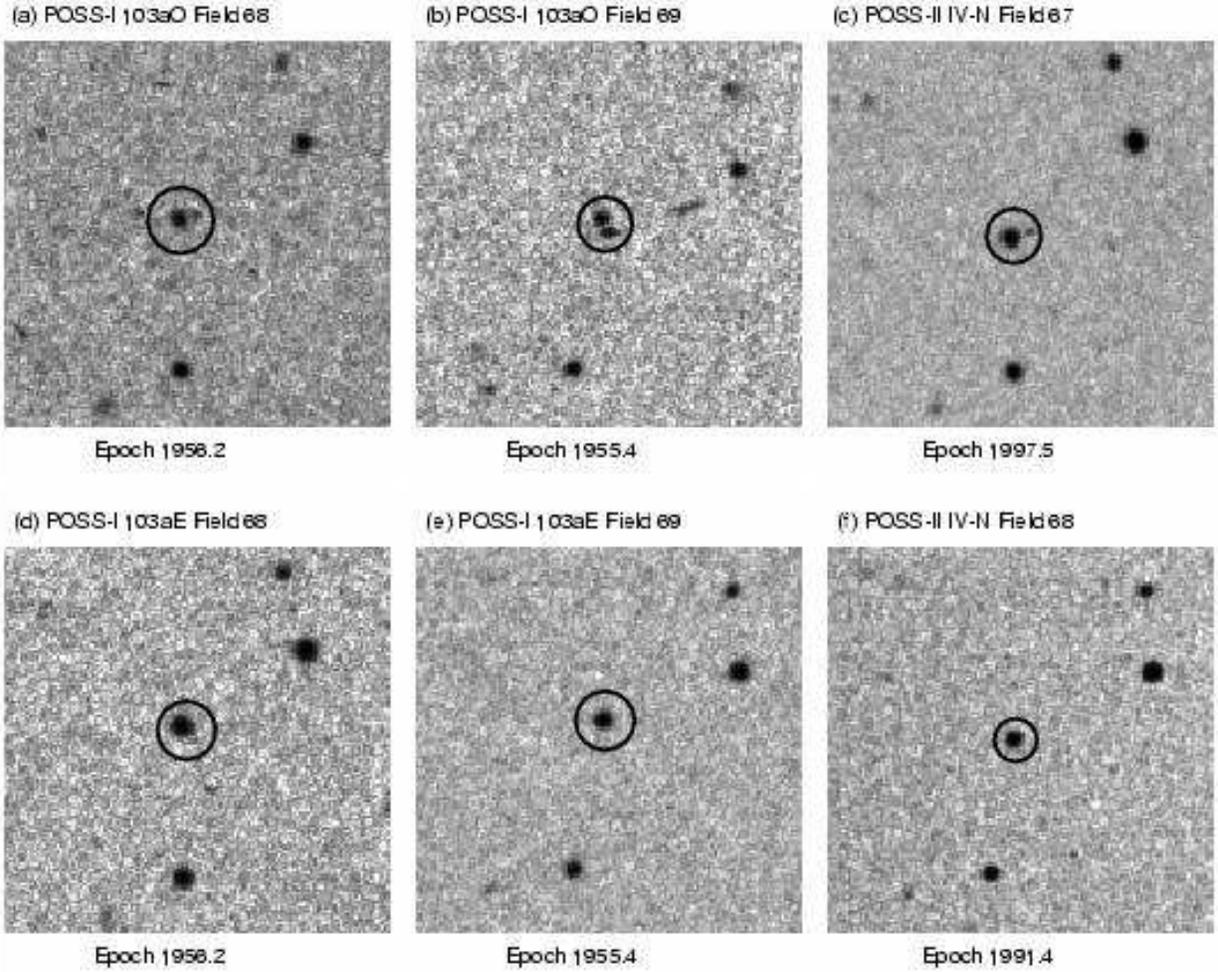}
\caption{The region around MUSR 40 (marked with a circle).  All images
  are $2{\arcmin}$ square.  Images (a)--(c) show more than one object
  within the circle around MUSR 40.  Images (d)--(f) show only MUSR 40
  within the circle.  Images (a) and (d) are from the POSS-I 103aO and
  103aE images of field 68 respectively. (b) and (e) are from the
  POSS-I 103aO and 103aE images of field 69 respectively. (c) and (f)
  are from the POSS-II IV-N images of fields 67 and 68 respectively.
  North is up, and East is to the right.
\label{fig-sr40}}
\end{figure}

\begin{figure}
\plotone{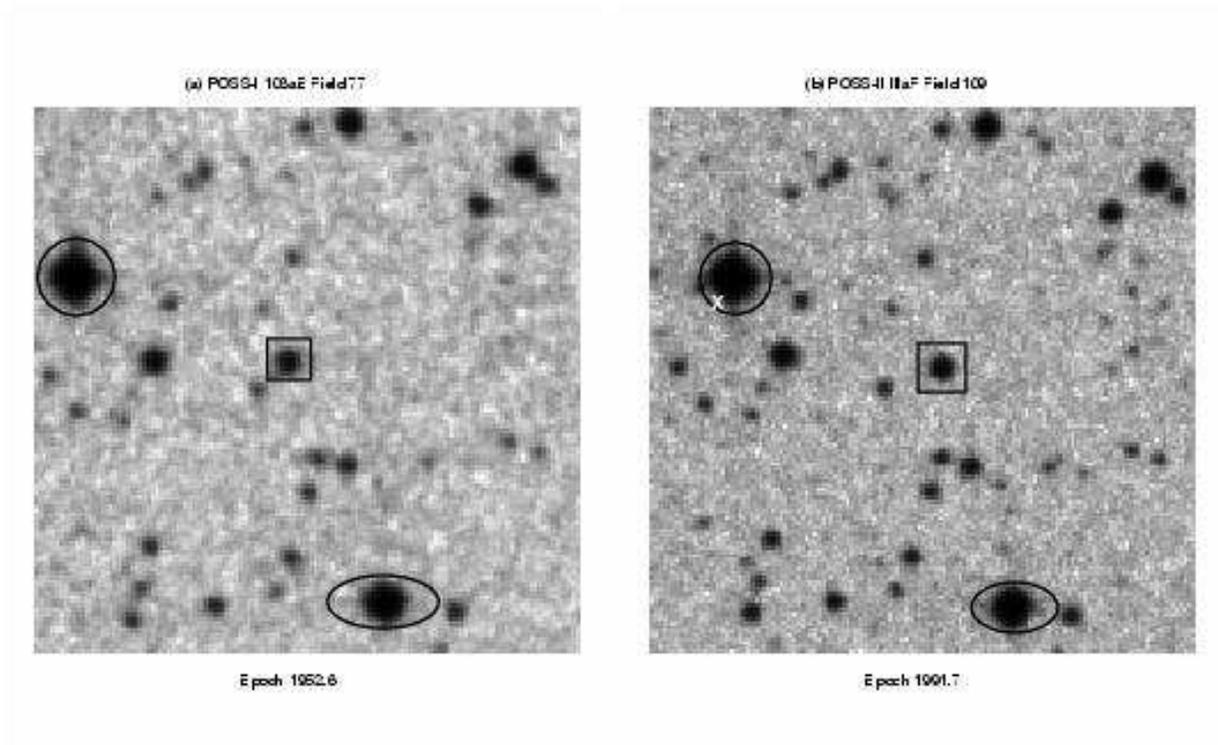}
\caption{The region around LDS~4990 (MUSR 56 - marked with a circle), USNO-B1
  1543-0282460 (MUSR 57 - marked with a square) and USNO-B1 1543-0282475 (MUSR 58
  - marked with an ellipse).  The position of 1RXS J224000.2+642310 is
  marked with a white X in (b).  Both images are $2{\arcmin}$ square.  (a) is
  from the POSS-I 103aE image of field 77. (b) is from the POSS-II
  IIIaF image of field 109.  North is up, and East is to the right.
\label{fig-lds4990}}
\end{figure}

\begin{figure}
\plotone{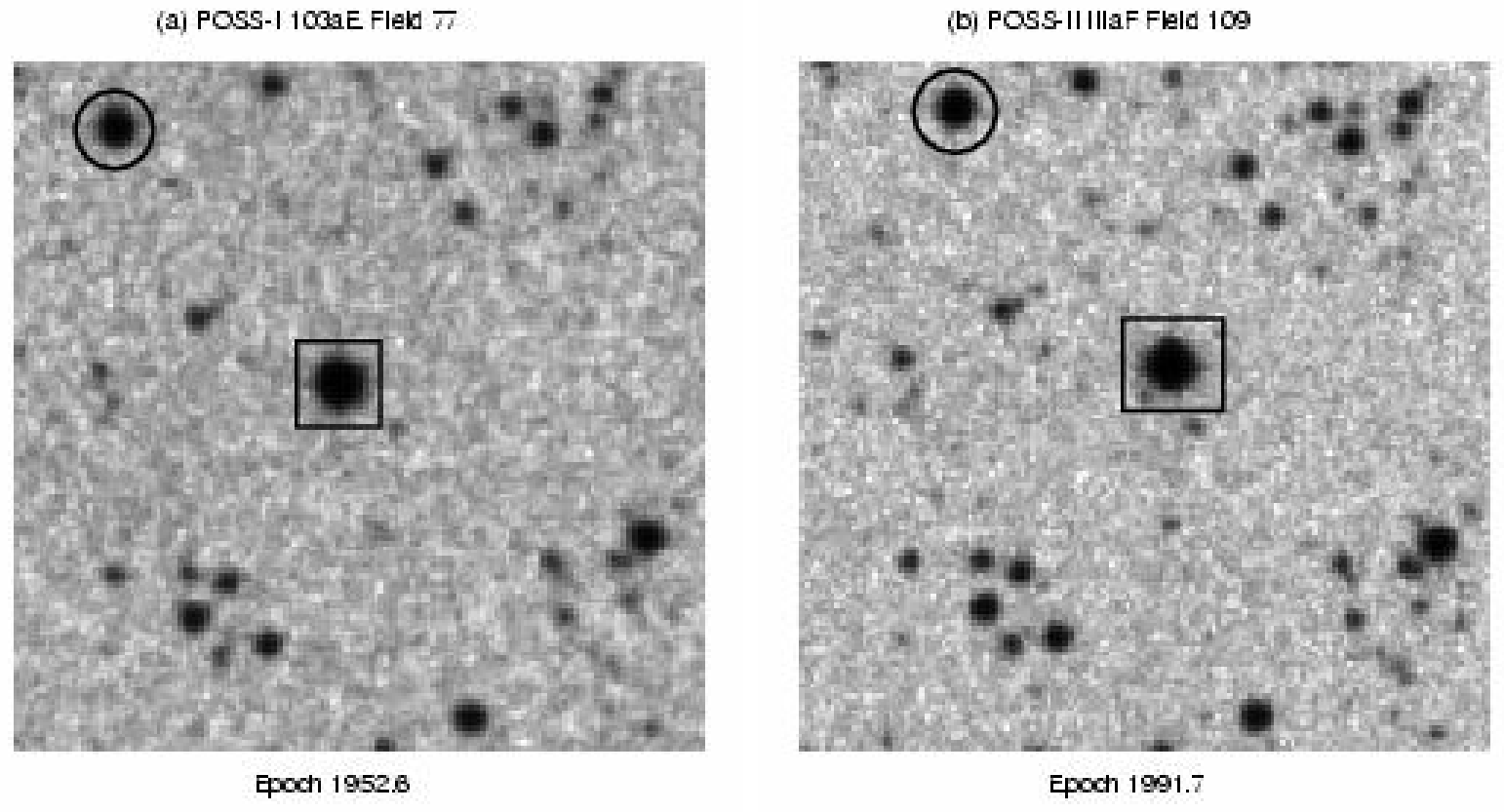}
\caption{The region around MUSR 61 (marked with a circle), and MUSR 62
  (marked with a square).  Both images are $2{\arcmin}$ square.  (a)
  is from the POSS-I 103aE image of field 77. (b) is from the POSS-II
  IIIaF image of field 109.  North is up, and East is to the right.
\label{fig-sr61_62}}
\end{figure}

\begin{figure}
\plotone{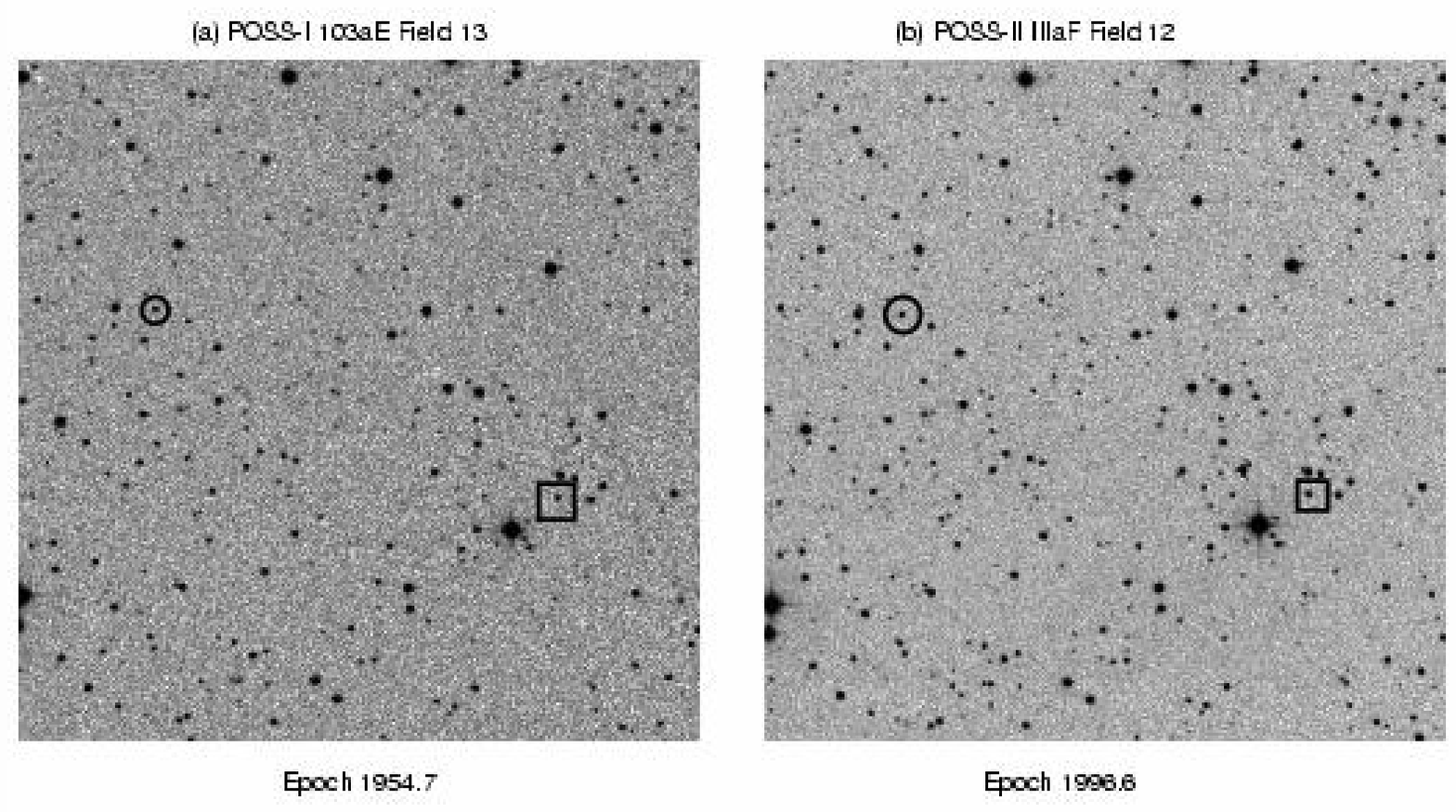}
\caption{The region around MUSR 76 (marked with a circle), and MUSR 77
  (marked with a square).  Both images are $10{\arcmin}$ square.  (a)
  is from the POSS-I 103aE image of field 13. (b) is from the POSS-II
  IIIaF image of field 12.  North is up, and East is to the right.
\label{fig-sr76_77}}
\end{figure}

\begin{figure}
\plottwo{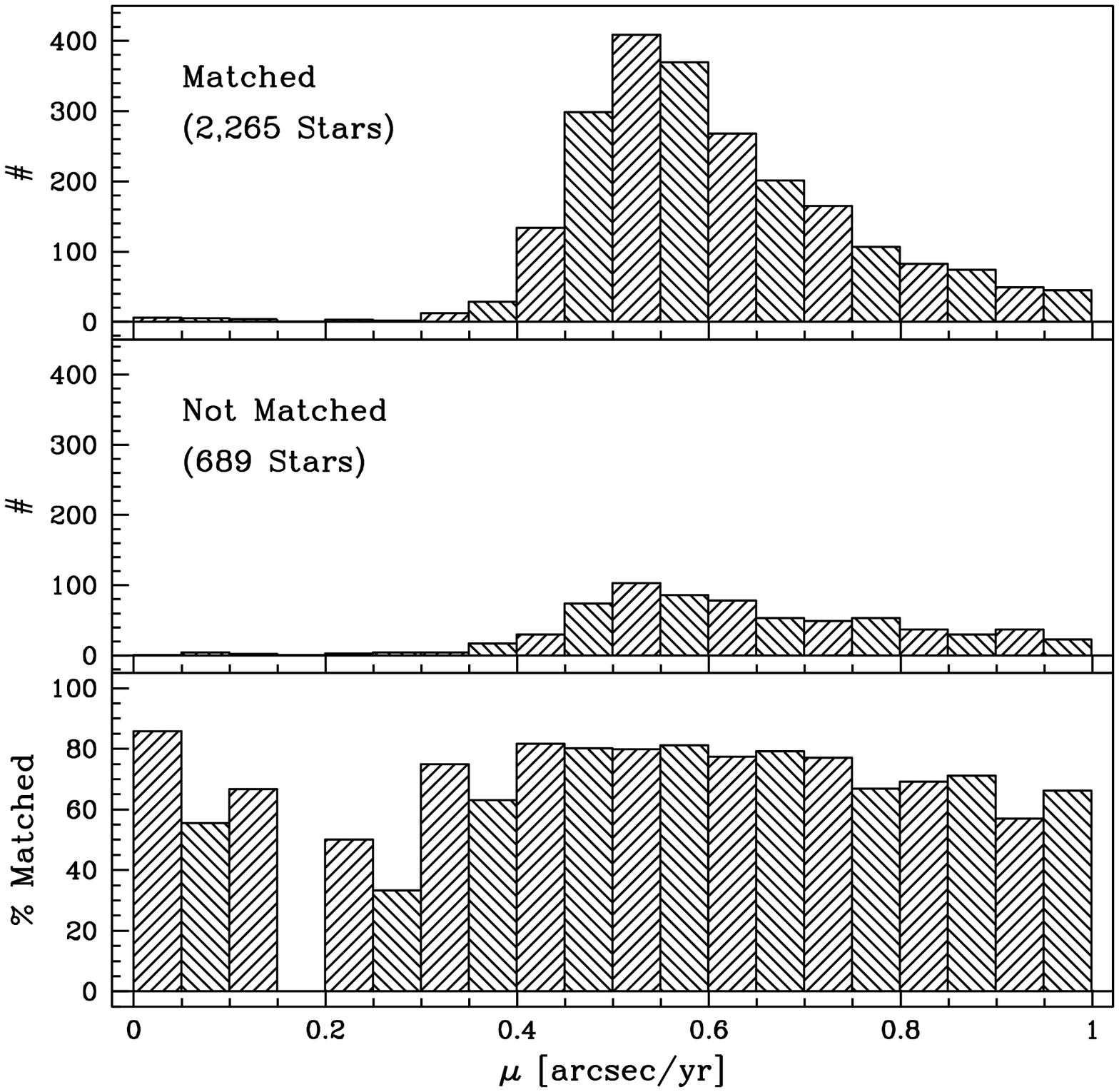}{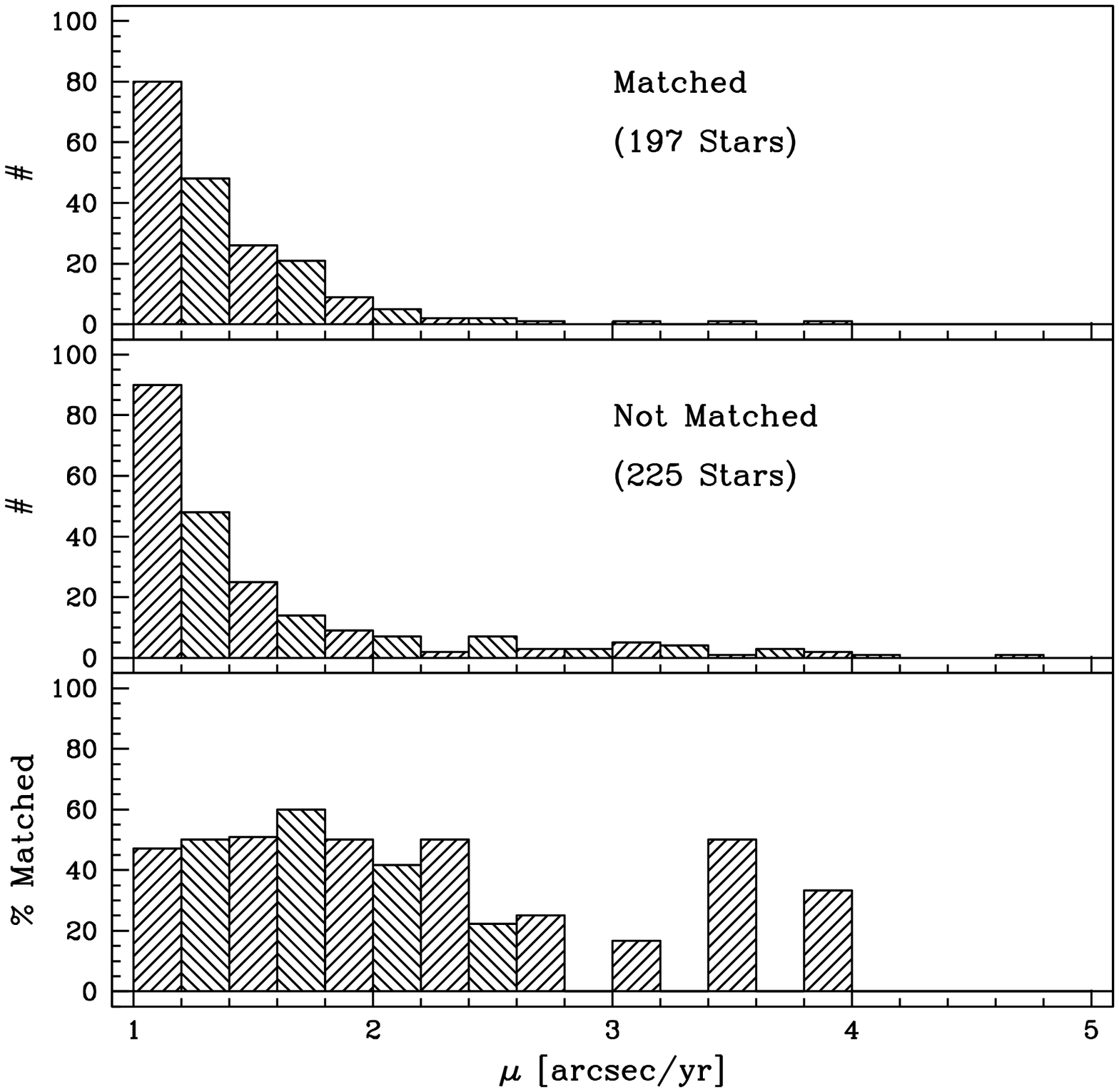}
\caption{Histograms of the number of rLHS objects matched (upper panels)
  and not matched (middle panels) with objects in USNO-B1.  The
  percent of objects matched is shown in the lower panels as a function
  of proper motion.  The left side shows the statistics for those objects
  with motions between 0 and 1 ${\rm arcsec \, yr^{-1}}$, and the
  right side shows the data for those objects with motions between 1
  and 5 ${\rm arcsec \, yr^{-1}}$.
\label{fig-ublhs-rat}}
\end{figure}

\begin{figure}
\plotone{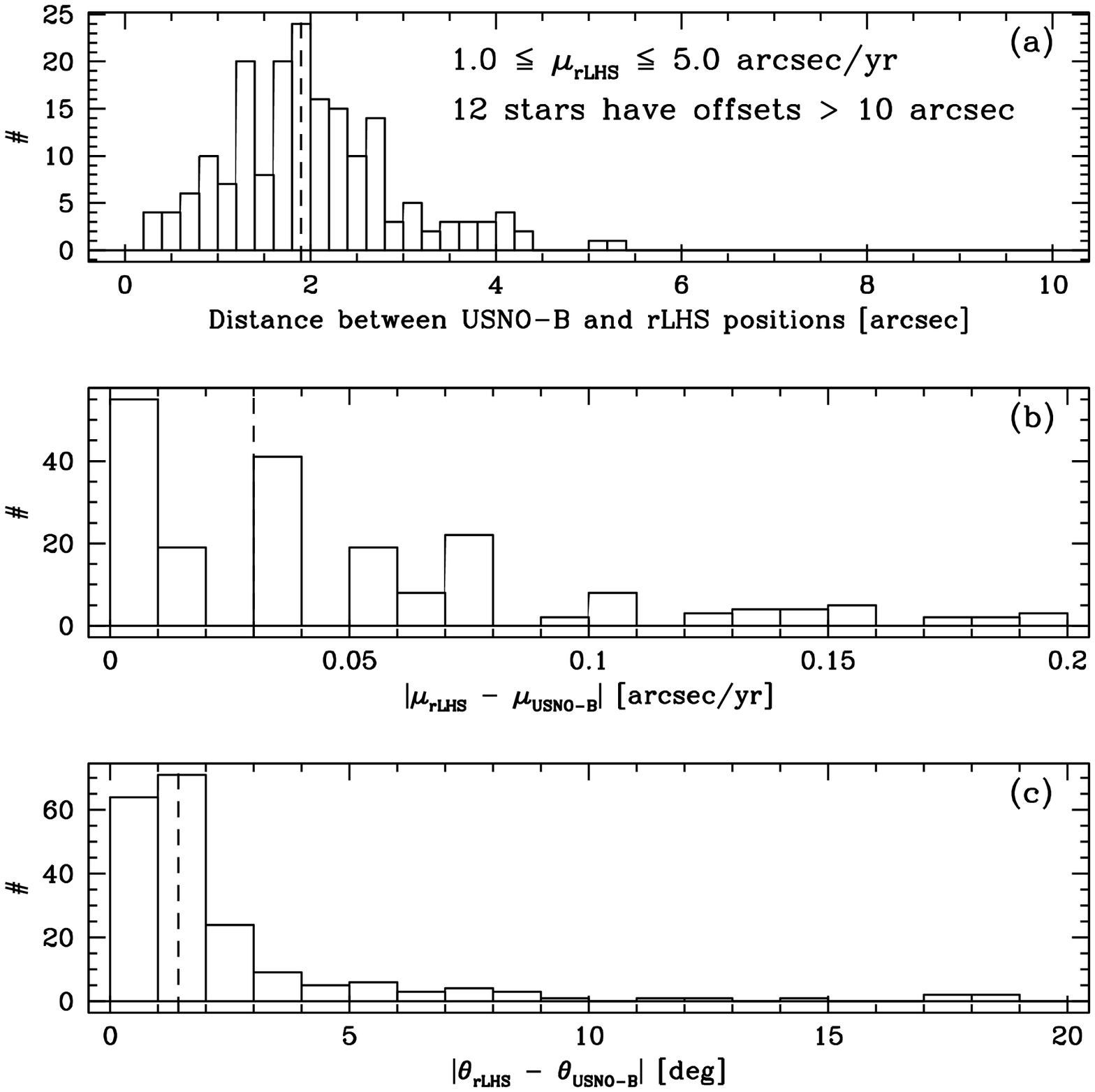}
\caption{Histograms showing the difference between rLHS objects that
  have been matched up to an object in USNO-B1.  Tycho-2 objects have
  been removed.  Panel (a) shows the mismatch in the position between
  USNO-B1 and rLHS for objects objects in USNO-B1 that matched high
  motion objects in the rLHS.  12 objects have position differences
  greater than 10 arcseconds.  Panel (b) shows the difference in
  magnitude of the proper motion, and (c) shows the difference in
  position angle.  These are effectively truncated at $\delta \mu <
  0.2 \, {\rm arcsec \, yr^{-1}}$ and $\delta\theta < 20{\degr}$ respectively
  by the initial matching search.  In all three panels, the median
  offset is marked by the dashed line.
\label{fig-ub_lhs}}
\end{figure}

Finally, It is important to know what the object density is like in
the region(s) you are interested in: if it is high (e.g. near the
galactic plane), then the contamination rate will rise as it becomes
progressively more difficult to unambiguously match up detections (see
Figs.~\ref{fig-one}).  J. Munn (private communication) noted during
the construction of the merged proper motion catalogue using USNO-B1
and SDSS DR1 data \citep{metal04} that requiring objects from USNO-B1
to have no neighbor within 7 arcseconds also helped to clean up the
contamination.

This particular work examining the high proper motion part of USNO-B1
has not made use of additional outside data.  As is clear from
\citet{metal04} and \citet{gk04}, it is possible to do a better job of
cleaning up the contamination in USNO-B1 if you have external data
with which to compare (e.g. the SDSS DR1 data).  If not, then you are
limited to methods similar to those used here, but it is fair to say
that the prospects of doing a decent job are still good.

\acknowledgments 

The author would like to thank D. Monet and all the staff at USNO for the
production of the USNO-B1 catalogue and all of the scans of the Schmidt
plates that have made this modest work possible.  I would like to thank
C. Dahn, and J. Munn for helpful discussions related to the USNO-B1
catalogue and high proper motion objects.  The author is grateful to the
referee for several helpful suggestions.

This research has made use of the USNOFS Image and Catalogue Archive
operated by the United States Naval Observatory, Flagstaff Station
(http://www.nofs.navy.mil/data/fchpix/).

The USNO-B1 catalogue contains data from a diverse collection of
photographs, reductions, and catalogues.  All the plates used were scanned
at the U. S. Naval Observatory Flagstaff Station, and the digitized
images are made available through their Image and Catalogue archive web
site.  A large number of different organizations claim copyright and/or
intellectual property rights on the various components.  This work is
based partly on photographic plates obtained at the Palomar Observatory
48 inch Oschin Schmidt Telescope for the POSS-I and POSS-II sky surveys.
The POSS-I was supported by grants from the National Geographic Society
and the California Institute of Technology.  The POSS-II was partially
supported by the Eastman Kodak Company, the National Geographic Society,
the Samuel Oschin Foundation, the Alfred P. Sloan Foundation, National
Science Foundation grants AST 84-08225, 87-19465, 90-23115 and 93-18984,
National Aeronautical and Space Administration grants NGL 05-002-140 and
NAGW-1710.

Some of the plates come from the UK Schmidt Telescope, which was
operated by the Royal Observatory Edinburgh, with funding from the UK
Science and Engineering Research Council, until June 1988, and
thereafter by the Anglo-Australian Observatory.  Additional plates
(ESO-R) come the European Southern Observatory, and were scanned with
the permission of ESO.

This research has made use of the SIMBAD database, and the Vizier
service operated at CDS, Strasbourg, France (http://cdsweb.u-strasbg.fr).

This publication makes use of data products from the Two Micron All Sky
Survey, which is a joint project of the University of Massachusetts and
the Infrared Processing and Analysis Center/California Institute of
Technology, funded by the National Aeronautics and Space Administration
and the National Science Foundation.

\clearpage

\begin{deluxetable}{llcccl}
\tablecaption{Photographic Source Material used in USNO-B1\tablenotemark{a}
 \label{tbl-plates}}
\tablewidth{0pt}
\tablehead{
\colhead{Survey} & 
\colhead{Emulsion} &
\colhead{Wavelength} & 
\colhead{Color\tablenotemark{b}} & 
\colhead{Declination\tablenotemark{c}} & 
\colhead{Epoch} \\
\colhead{} & 
\colhead{} &
\colhead{[nm]} &
\colhead{} & 
\colhead{[deg]} & 
\colhead{}
}
\startdata
POSS-I  & 103aO & 350--500 & B & $-30$ to $+90$ & 1949--1965 \\
POSS-I  & 103aE & 620--670 & R & $-30$ to $+90$ & 1949--1965 \\
POSS-II & IIIaJ & 385--540 & B & {\phs}{\phn}$0$ to $+90$   & 1985--2000 \\
POSS-II & IIIaF & 610--690 & R & {\phs}{\phn}$0$ to $+90$   & 1985--2000 \\
POSS-II & IV-N  & 730--900 & I & {\phs}{\phn}$0$ to $+90$   & 1989--2000 \\
SERC-J  & IIIaJ & 395--540 & B & $-90$ to $-20$ & 1978--1990 \\
SERC-EJ & IIIaJ & 395--540 & B & $-15$ to {\phn}$-5$  & 1984--1998 \\
ESO-R   & IIIaF & 630--690 & R & $-90$ to $-35$ & 1974--1987 \\
AAO-R   & IIIaF & 590--690 & R & $-90$ to $-20$ & 1985--1998 \\
SERC-ER & IIIaF & 590--690 & R & $-15$ to {\phn}$-5$  & 1979--1994 \\
SERC-I  & IV-N  & 715--900 & I & $-90$ to {\phs}{\phn}$0$   & 1978--2002 \\
SERC-I\tablenotemark{d}  & IV-N  & 715--900 & I & {\phn}$+5$ to $+20$  & 1981--2002 \\
\enddata
\tablenotetext{a}{The contents of this table follow from
  \citet{metal03} Table 1.}
\tablenotetext{b}{The colors listed here are rough PHOTOGRAPHIC
  colors.  They correspond to the magnitudes given in USNO-B1.}
\tablenotetext{c}{The range in declination of the field centers in
  each survey used in the construction of USNO-B1.}
\tablenotetext{d}{These fields are an extension of the SERC-I that was
  done to fill in fields that were not taken during the POSS-II IV-N survey.}
\end{deluxetable}

\clearpage

\begin{deluxetable}{ccccccccccl}
\tabletypesize{\scriptsize}
\tablecaption{Known, Unflagged Objects with Motions Between 1.0 and
 5.0 $\arcsec \, {\rm yr^{-1}}$. \label{tbl-known}}
\tablewidth{0pt}
\tablecolumns{11}
\tablehead{
\colhead{USNO-B} & 
\colhead{$\mu$\tablenotemark{a}} &
\colhead{$\theta$\tablenotemark{a}} & 
\colhead{RA} & 
\colhead{Dec} &
\colhead{{\it l}} &
\colhead{{\it b}} & 
\colhead{$\rm B_2$\tablenotemark{b}} & 
\colhead{$\rm R_2$\tablenotemark{b}} & 
\colhead{$\rm I_2$\tablenotemark{b}} & 
\colhead{AltID\tablenotemark{c}} \\
\colhead{ID} & 
\colhead{[${\arcsec} \, {\rm yr^{-1}}$]} &
\colhead{[deg]} & 
\colhead{[hrs]} & 
\colhead{[deg]} &
\colhead{[deg]} &
\colhead{[deg]} &
\colhead{[mag]} & 
\colhead{[mag]} & 
\colhead{[mag]} & 
\colhead{}
}
\startdata
\cutinhead{LHS Objects\tablenotemark{c}}
0121-0045493 & 1.03 & 142.7 & 09.28481 & $-77.8234$ & 292.5559 & $-$19.5296      & 14.38 & 12.11 & 10.14 & 263 \\
0185-0249424 & 1.16 & 338.8 & 12.47781 & $-71.4644$ & 301.1041 & {\phn}$-$8.6735 & 15.15 & 12.72 & 11.10 & 328 \\
0185-0249438 & 1.18 & 338.9 & 12.47866 & $-71.4656$ & 301.1083 & {\phn}$-$8.6743 & 16.48 & 14.72 & 12.19 & 329 \\
0222-0190851 & 2.14 & 136.6 & 07.88561 & $-67.7924$ & 280.2038 & $-$19.4322      & 14.53 & 13.77 & 13.02 & 34  \\
0289-0005722 & 1.11 & {\phn}93.7 & 00.82473 & $-61.0424$ & 303.3558 & $-$56.0842      & 13.03 & 11.50 & 10.05 & 124 \\
0560-0118956 & 1.67 & 351.6 & 07.76069 & $-33.9311$ & 248.9038 & $-$4.6752 & 17.40 & 15.67 & 15.60 & 237a \\
1611-0086923 & 1.91 & 256.0 & 10.61735 & $+71.1830$ & 136.8566 & {\phs}42.1223   & 17.31 & 16.14 & 15.47 & 285 \\ 
1688-0078160 & 1.16 & {\phn}63.0 & 21.68144 & $+78.8227$ & 114.2010 & {\phs}19.3348   & 14.77 & 12.91 & 10.89 & 514 \\
1695-0027702 & 1.20 & 136.2 & 05.63677 & $+79.5221$ & 133.7772 & {\phs}23.4244   & 19.93 & 17.26 & 13.77 & 207 \\
\cutinhead{LSR Objects\tablenotemark{c}}
0872-0489450 & 1.01 & 214.9 & 18.16393 & $-02.7953$ & {\phn}25.6449 & {\phn}7.9460 & 17.29 & 15.42 & 13.03 & $1809-0247$ \\
1042-0321115 & 1.00 & 235.4 & 17.97303 & $+14.2939$ & {\phn}40.0841 &      18.0868 & 17.24 & 16.35 & 15.57 & $1758+1417$ \\
1068-0333681 & 1.00 & 117.0 & 17.92576 & $+16.8164$ & {\phn}42.2300 &      19.7325 & 17.00 & 14.84 & 12.49 & $1755+1648$ \\
1207-0075220 & 1.10 & 148.5 & 05.08660 & $+30.7256$ & 173.6030      &      -6.2643 & 18.12 & 16.33 & 14.73 & $0505+3043$ \\
1325-0110870 & 1.54 & 159.6 & 04.33114 & $+42.5585$ & 158.6564      &      -5.4068 & 20.26 & 17.35 & 14.43 & $0419+4233$ \\
1491-0005115 & 1.47 & 217.9 & 00.19217 & $+59.1445$ & 117.8254      &      -3.3310 & 16.70 & 14.85 & 11.38 & $0011+5908$ \\
1491-0151160 & 1.01 & 173.5 & 05.25859 & $+59.1883$ & 151.5063      &      11.9084 & 19.97 & 16.60 & 14.35 & $0515+5911$ \\
\cutinhead{Assorted Objects}
0143-0198407 & 1.04 & 143.8 & 21.25418 & $-75.6977$ & 317.0296           & $-$34.8155      & 16.13 & 13.44 & 11.24 & SC,P \\  
0258-0023144 & 1.06 & 140.9 & 03.71595 & $-64.1322$ & 278.5502           & $-$44.0139      & 17.15 & 15.04 & 12.66 & SC \\  
0279-0008695 & 1.10 &  82.3 & 00.87091 & $-62.0317$ & 302.7644           & $-$55.0962      & 19.65 & 16.72 & 13.33 & SIPS  \\  
0300-0785973 & 1.42 & 165.4 & 20.20883 & $-59.9476$ & 337.1327           & $-$33.3064      & 16.63 & 15.37 & 14.86 & P \\  
0358-0039309 & 1.07 & 168.6 & 05.00438 & $-54.1077$ & 261.9192           & $-$37.7847      & 19.88 & 17.61 & 15.45 & P \\  
0393-0108806 & 1.00 & 326.5 & 08.50019 & $-50.6624$ & 267.4696           & {\phn}$-$6.7440 & 15.67 & 13.76 & 12.52 & L \\  
0443-0286531 & 1.31 & 281.9 & 12.46300 & $-45.6879$ & 298.6155           & {\phs}16.9854   & 16.37 & 14.49 & 13.16 & L \\ 
0477-0913359 & 1.03 & 171.7 & 19.94933 & $-42.2729$ & 357.5855           & $-$29.5013      & 18.72 & 17.03 & 13.71 & R,P \\  
0500-0227632 & 1.52 & 229.4 & 10.80405 & $-39.9353$ & 278.6839           & {\phs}17.0658   & 18.58 & 15.93 & 12.66 & D \\ 
0510-0792885 & 1.07 & 109.6 & 22.24298 & $-38.9852$ & {\phn}{\phn}2.7952 & $-$55.3720      & 16.39 & 15.05 & 15.09 & P, O\\  
0533-0785516 & 1.29 & 184.6 & 19.27960 & $-36.6349$ & {\phn}{\phn}1.3299 & $-$20.5417      & 18.08 & 15.76 & 14.85 & L \\ 
0847-0018930 & 1.04 &  67.3 & 02.08655 & $-05.2983$ & 165.0326           & $-$61.9816      & 18.86 & 17.86 & 17.24 & O \\  
\enddata
\tablenotetext{a}{The proper motions are relative to the reference
frame established by the YS4.0 catalogue stars (see Monet et al. 2003
for details).}
\tablenotetext{b}{Photographic magnitudes from the second epoch
Schmidt surveys (POSS-II in the north, SERC-J, SERC-EJ, AAO-R,
SERC-ER, and SERC-I in the south).}
\tablenotetext{c}{References: For the LHS objects, the AltID is the
  LHS \citep{lhs} catalogue number.  For the LSR objects, the AltID is
  the id given in \citet{lsr02}.  For the assorted objects their
  source is given by this key: D=\citet{detal01}, L=\citet{l05},
  O=\citet{oetal01}, P=\citet{petal03,petal04}, R=\citet{rrsi02},
  SC=\citet{hetal04}, SIPS=\citet{dhc05}}
\end{deluxetable}

\begin{deluxetable}{cccccccccccccrl}
\rotate
\tabletypesize{\scriptsize}
\tablecaption{Objects with Motions less than 1.0
  ${\arcsec} \, {\rm yr^{-1}}$. \label{tbl-ltone}}
\tablewidth{0pt}
\tablehead{
\colhead{USNO-B} & 
\colhead{$\mu$\tablenotemark{a}} &
\colhead{$\theta$\tablenotemark{a}} & 
\colhead{RA} & 
\colhead{Dec} &
\colhead{{\it l}} &
\colhead{{\it b}} &
\colhead{$\rm B_2$\tablenotemark{b}} & 
\colhead{$\rm R_2$\tablenotemark{b}} & 
\colhead{$\rm I_2$\tablenotemark{b}} & 
\colhead{$\rm J$\tablenotemark{c}} & 
\colhead{$\rm H$\tablenotemark{c}} & 
\colhead{$\rm K_s$\tablenotemark{c}} &
\colhead{Class\tablenotemark{d}} &
\colhead{AltID\tablenotemark{e}} \\
\colhead{ID} & 
\colhead{[${\arcsec} \, {\rm yr^{-1}}$]} &
\colhead{[deg]} & 
\colhead{[hrs]} & 
\colhead{[deg]} &
\colhead{[deg]} &
\colhead{[deg]} &
\colhead{[mag]} & 
\colhead{[mag]} & 
\colhead{[mag]} & 
\colhead{[mag]} & 
\colhead{[mag]} & 
\colhead{[mag]} &
\colhead{} &
\colhead{}
}
\startdata
0338-0848607 & 0.31 & {\phn}71.8 & 23.41315 & $-56.1517$ & 325.1606      & -57.0780 & 12.36\tablenotemark{f} & 10.41 & 10.06 & {\phn}9.36 & {\phn}8.74 & {\phn}8.59   & d & NLTT 9526 \\
1180-0331814 & 0.09 & 210.6      & 18.08276 & $+28.0142$ & {\phn}54.2204 & {\phs}21.8639 & 14.93 & 14.45 & 14.05 & 13.46 & 13.24 & 13.12 & sd &  \\  
1686-0094267 & 0.40 & {\phn}75.0 & 23.96100 & $+78.6681$ & 120.2067      & {\phs}16.0819 & 17.64 & 17.46 & 17.55 & 16.31 & 15.49 & 15.68 & wd &  \\
1698-0001063 & 0.09 & {\phn}63.0 & 00.23815 & $+79.8183$ & 121.2218      & {\phs}17.0701 & 19.18 & 17.29 & 16.46 & 14.89 & 14.30 & 14.05 & d &  \\
\enddata
\tablenotetext{a}{The proper motions are relative to the reference
  frame established by the YS4.0 catalogue stars (see Monet et al. 2003
  for details).}
\tablenotetext{b}{Photographic magnitudes from the second epoch
  Schmidt surveys (POSS-II in the north, SERC-J, SERC-EJ, AAO-R,
  SERC-ER, and SERC-I in the south).}
\tablenotetext{c}{Near IR magnitudes are from the 2MASS final release
point source catalogue \citep{2mass}.}
\tablenotetext{d}{Classification: d = dwarf, sd = sub-dwarf, wd =
  white dwarf.}
\tablenotetext{e}{Alternate Identification: NLTT = \citet{nltt}}
\tablenotetext{f}{Magnitude taken from another USNO-B1 catalogue entry,
  which was made up of additional detections of this objects.}
\end{deluxetable}

\begin{deluxetable}{cccccccccccccr}
\rotate
\tabletypesize{\scriptsize}
\tablecaption{New Objects with Motions Between 1.0 and 5.0
  ${\arcsec}\, {\rm yr^{-1}}$ . \label{tbl-new}} 
\tablewidth{0pt}
\tablehead{
\colhead{USNO-B} & 
\colhead{$\mu$\tablenotemark{a}} &
\colhead{$\theta$\tablenotemark{a}} & 
\colhead{RA} & 
\colhead{Dec} &
\colhead{{\it l}} &
\colhead{{\it b}} &
\colhead{$\rm B_2$\tablenotemark{b}} & 
\colhead{$\rm R_2$\tablenotemark{b}} & 
\colhead{$\rm I_2$\tablenotemark{b}} & 
\colhead{$\rm J$\tablenotemark{c}} & 
\colhead{$\rm H$\tablenotemark{c}} & 
\colhead{$\rm K_s$\tablenotemark{c}} &
\colhead{Class\tablenotemark{d}} \\
\colhead{ID} & 
\colhead{[${\arcsec}\, {\rm yr^{-1}}$]} &
\colhead{[deg]} & 
\colhead{[hrs]} & 
\colhead{[deg]} &
\colhead{[deg]} &
\colhead{[deg]} &
\colhead{[mag]} & 
\colhead{[mag]} & 
\colhead{[mag]} & 
\colhead{[mag]} & 
\colhead{[mag]} & 
\colhead{[mag]} &
\colhead{}
}
\startdata
0484-0243338 & 1.20 & 282.6 & 11.13220 & $-41.5980$ & 282.9517 & 17.2231   & 16.88 & 14.27 & 13.04 & 12.19 & 11.69 & 11.47 & sd  \\
0867-0249298 & 1.08 & 226.9 & 11.62128 & $-03.2934$ & 269.6031 & 54.7070   & 16.23 & 14.12 & 12.35 & 10.87 & 10.36 & 10.09 & d \\
\enddata
\tablenotetext{a}{The proper motions are relative to the reference
  frame established by the YS4.0 catalogue stars (see Monet et al. 2003
  for details).}
\tablenotetext{b}{Photographic magnitudes from the second epoch
  Schmidt surveys (POSS-II in the north, SERC-J, SERC-EJ, AAO-R,
  SERC-ER, and SERC-I in the south).}
\tablenotetext{c}{Near IR magnitudes are from the 2MASS final release
  point source catalogue \citep{2mass}.}
\tablenotetext{d}{Classification: d = dwarf, sd = sub-dwarf, wd =
  white dwarf.}
\end{deluxetable}

\clearpage

\begin{deluxetable}{lrrrrrrrrrrrrrcl}
\rotate
\tabletypesize{\scriptsize}
\tablecaption{Serendipitous objects with good solutions in USNO-B1. \label{tbl-serub}}
\tablewidth{0pt}
\tablehead{
\colhead{ID\tablenotemark{a}} & 
\colhead{$\mu$\tablenotemark{b}} &
\colhead{$\theta$\tablenotemark{b}} & 
\colhead{RA} & 
\colhead{Dec} &
\colhead{{\it l}} &
\colhead{{\it b}} &
\colhead{$\rm B$\tablenotemark{c}} & 
\colhead{$\rm R$\tablenotemark{c}} & 
\colhead{$\rm I$\tablenotemark{c}} & 
\colhead{$\rm J$\tablenotemark{d}} & 
\colhead{$\rm H$\tablenotemark{d}} & 
\colhead{$\rm K_s$\tablenotemark{d}} & 
\colhead{\#\tablenotemark{e}} &
\colhead{Class\tablenotemark{f}} &
\colhead{AltID\tablenotemark{g}} \\
\colhead{} & 
\colhead{[${\rm mas \, yr^{-1}}$]} &
\colhead{[deg]} & 
\colhead{[hrs]} & 
\colhead{[deg]} &
\colhead{[deg]} &
\colhead{[deg]} &
\colhead{[mag]} & 
\colhead{[mag]} & 
\colhead{[mag]} & 
\colhead{[mag]} & 
\colhead{[mag]} & 
\colhead{[mag]} &
\colhead{} & 
\colhead{} &
\colhead{} \\
\colhead{(1)} & 
\colhead{(2)} &
\colhead{(3)} & 
\colhead{(4)} & 
\colhead{(5)} &
\colhead{(6)} &
\colhead{(7)} & 
\colhead{(8)} &
\colhead{(9)} &
\colhead{(10)} &
\colhead{(11)} &
\colhead{(12)} & 
\colhead{(13)} & 
\colhead{(14)} & 
\colhead{(15)} &
\colhead{(16)}
}
\startdata
4492-01044-1 &  225.6 &   65.9 &  0.33483 &  76.1291 & 121.0064 &  13.3746 & 12.7 & 11.4 & 10.9 &  9.2 &  8.6 &  8.4 &  0 & d  & TYC2-4492-1044-1 \\
1660-0002691 &   78.2 &   94.4 &  0.34624 &  76.0645 & 121.0401 &  13.3055 & 18.2 & 16.2 & 15.0 & 13.4 & 12.8 & 12.5 &  5 & d  &  \\
1662-0002920 &   87.3 &  110.1 &  0.37893 &  76.2253 & 121.1795 &  13.4513 & 17.5 & 15.8 & 14.2 & 12.1 & 11.5 & 11.3 &  5 & d  &  \\
1657-0005791 &  164.1 &    9.1 &  0.49845 &  75.7001 & 121.5699 &  12.8853 & 16.0 & 13.9 & 12.7 & 10.8 & 10.3 & 10.0 &  5 & d  &  \\
1698-0004712 &   81.2 &   99.9 &  0.92103 &  79.8888 & 123.1074 &  17.0183 & 18.1 & 16.3 & 15.8 & 14.3 & 13.7 & 13.4 &  5 & d  &  \\
1695-0005846 &  373.4 &   69.6 &  1.00922 &  79.5763 & 123.3623 &  16.7121 & 17.4 & 15.2 & 12.8 & 11.7 & 11.1 & 10.8 &  5 & d  & NLTT-3242 \\
0855-0009685 &  189.7 &   65.1 &  1.01479 &  -4.4809 & 129.0295 & -67.2407 & 14.7 & 12.8 & 11.8 & 10.7 & 10.1 &  9.9 &  5 & d  & UCAC2-30296684 \\
0855-0009698 & 1321.5 &   70.3 &  1.01566 &  -4.4490 & 129.0551 & -67.2077 & 14.4 & 12.3 & 10.4 &  9.0 &  8.5 &  8.2 &  5 & d  & LHS-130 \\
1540-0035963 &  128.6 &    5.4 &  1.17036 &  64.0343 & 124.9866 &   1.2375 & 17.2 & 16.8 & 17.2 & 16.1 & 15.5 & 15.1 &  5 & sd &  \\
1407-0071339 &  120.9 &  145.8 &  2.66405 &  50.7820 & 139.8638 &  -8.4912 & 15.4 & 12.7\tablenotemark{h} & 11.4 & 10.7 & 10.1 &  9.9 &  4 & d &  \\
1683-0025701 &   93.3 &  135.0 &  3.73783 &  78.3068 & 131.3389 &  18.3639 & 15.5 & 14.7 & 13.9 & 12.4 & 11.8 & 11.7 &  5 & d  &  \\
1544-0112254 &   84.4 &  148.6 &  4.12699 &  64.4990 & 142.1650 &   9.2308 & 20.2 & 17.4 & 16.6 & 15.2 & 14.4 & 14.3 &  5 & d  &  \\
1522-0148544 &  113.6 &  118.4 &  4.32810 &  62.2920 & 144.7008 &   8.5851 & 19.8 & 17.5 & 16.4 & 15.2 & 14.5 & 14.4 &  5 & sd &  \\
1574-0111126 &   80.6 &  156.6 &  5.51495 &  67.4569 & 145.1288 &  17.6793 & 17.7 & 15.8 & 14.6 & 13.5 & 12.8 & 12.6 &  4 & d  &  \\
1578-0121823 &  235.8 &  169.7 &  6.37029 &  67.8010 & 146.8259 &  22.2983 & 15.9 & 14.1 & 11.4 & 10.7 & 10.2 &  9.8 &  5 & d  &  \\
1659-0050193 &  324.0 &  173.6 &  7.48851 &  75.9009 & 138.8070 &  28.5491 & 19.1 & 16.6 & 14.2 & 12.0 & 11.5 & 11.1 &  5 & d  & NLTT-17835 \\
4133-00625-1 &  502.0 &  180.5 &  8.42792 &  66.4623 & 149.1560 &  34.0631 &  9.3 &  8.3 &  7.8 &  7.2 &  6.8 &  6.7 &  0 & d  & TYC2-4133-00625-1 \\
1575-0148828 &   26.9 &  132.0 & 10.29553 &  67.5945 & 141.9665 &  43.4178 & 14.2 & 12.5 & 11.2 & 10.8 & 10.2 & 10.1 &  5 & d  &  \\
1576-0150230 &   58.1 &  229.2 & 10.29363 &  67.6537 & 141.9141 &  43.3712 & 13.2 & 11.7 & 11.0 & 10.5 & 10.0 &  9.9 &  5 & d  &  \\
0867-0255298 &   58.5 &  262.1 & 11.95509 &  -3.2049 & 277.5471 &  56.9749 & 16.6 & 15.6 & 14.2 & 13.5 & 12.8 & 12.6 &  5 & d  &  \\
0867-0255338 &   57.3 &  282.1 & 11.95771 &  -3.2520 & 277.6467 &  56.9463 & 15.3 & 14.2 & 14.2 & 13.6 & 13.1 & 13.0 &  5 & sd &  \\
1662-0061497 &   84.3 &  292.3 & 12.50206 &  76.2446 & 124.6039 &  40.8144 & 19.7 & 17.2 & 16.0 & 14.3 & 13.8 & 13.6 &  5 & d  &  \\
1663-0069093 &   76.1 &  273.0 & 15.09013 &  76.3685 & 113.4095 &  38.1747 & 14.8 & 14.3 & 14.0 & 13.2 & 12.9 & 12.9 &  5 & sd &  \\
1570-0182321 &  101.1 &  245.5 & 16.58837 &  67.0294 &  98.7379 &  37.8952 & 17.2 & 15.3 & 12.9 & 11.9 & 11.3 & 11.1 &  5 & d  &  \\
1548-0247178 &  273.9 &   83.3 & 21.69641 &  64.8562 & 104.4654 &   9.0259 & 17.9 & 16.1 & 13.3 & 12.0 & 11.4 & 11.1 &  5 & d  & NLTT-51912 \\
1371-0540717 &   90.0 &   90.0 & 22.12871 &  47.1923 &  96.2810 &  -7.0307 & 17.5 & 15.1 & 13.7 & 13.4 & 12.8 & 12.7 &  5 & d  &  \\
1503-0343417 &   74.7 &  195.5 & 22.52417 &  60.3793 & 106.4534 &   2.0660 & 18.7 & 16.2 & 14.5 & 13.3 & 12.8 & 12.6 &  5 & d  &  \\
1543-0282460 &  152.3 &  103.7 & 22.66878 &  64.3816 & 109.3223 &   5.0372 & 17.9 & 15.8 & 14.2 & 12.4 & 11.8 & 11.5 &  5 & d  & LDS-4990 \\
1543-0282475 &   55.7 &  111.0 & 22.66946 &  64.3670 & 109.3190 &   5.0222 & 15.4 & 13.4 & 12.7 & 11.2 & 10.5 & 10.4 &  5 & d  &  \\
1680-0117205 &  106.0 &   54.2 & 22.73451 &  78.0837 & 116.3959 &  16.8607 & 17.0 & 15.4 & 14.3 & 13.1 & 12.6 & 12.3 &  5 & d  &  \\
1544-0281760 &   86.3 &  166.6 & 22.76294 &  64.4808 & 109.9080 &   4.8349 & 19.4 & 17.2 & 16.2 & 14.5 & 13.9 & 13.6 &  5 & d  &  \\
1543-0289304 &   48.7 &   70.8 & 22.92054 &  64.3196 & 110.7505 &   4.2318 & 18.1 & 15.8 & 14.8 & 13.2 & 12.6 & 12.3 &  5 & d  &  \\
1558-0247908 &   69.4 &   41.5 & 23.07854 &  65.8848 & 112.3157 &   5.2374 & 17.4 & 15.2 & 14.7 & 13.7 & 13.0 & 12.8 &  5 & d  &  \\
1558-0247969 &  116.3 &  310.8 & 23.08128 &  65.8882 & 112.3325 &   5.2338 & 20.6 & 18.3\tablenotemark{h} & 16.5 & 14.8 & 14.3 & 14.1 &  4 & d  &  \\
1559-0249177 &  139.7 &   99.1 & 23.14135 &  65.9357 & 112.6906 &   5.1324 & 17.8 & 15.4 & 14.6 & 12.7 & 12.0 & 11.8 &  4 & d  &  \\
1691-0087337 &  154.1 &   92.2 & 23.36841 &  79.1872 & 118.6554 &  17.0646 & 16.8 & 15.4 & 13.8 & 12.4 & 11.8 & 11.6 &  5 & d  &  \\
1675-0139929 &  157.5 &   82.0 & 23.59605 &  77.5049 & 118.7623 &  15.2500 & 17.4 & 15.4 & 14.2 & 12.6 & 12.1 & 11.8 &  5 & d  &  \\
1675-0140048 &  103.7 &  129.5 & 23.60936 &  77.5253 & 118.8113 &  15.2566 & 16.8 & 15.0 & 14.0 & 12.8 & 12.2 & 12.0 &  5 & d  &  \\
1675-0140100 &  109.3 &   34.6 & 23.61504 &  77.5031 & 118.8228 &  15.2300 & 17.4 & 16.1 & 14.2 & 12.4 & 11.8 & 11.5 &  5 & d  &  \\
1675-0140133 &  434.8 &   93.4 & 23.61889 &  77.5720 & 118.8562 &  15.2922 & 14.7 & 12.9 & 10.8 & 10.2 &  9.7 &  9.4 &  5 & d  & NLTT-57436 \\
1663-0113155 &   71.0 &   80.3 & 23.82149 &  76.3079 & 119.1870 &  13.8901 & 17.9 & 15.8 & 15.2 & 14.1 & 13.3 & 13.3 &  5 & d  &  \\
\enddata
\tablenotetext{a}{IDs of the form ZZZZ-NNNNNNN are from USNO-B1, and those
  of the form ZZZZ-RRRRR-N are from Tycho-2.}
\tablenotetext{b}{The proper motions are relative to the reference
  frame established by the YS4.0 catalogue stars (see Monet et al. 2003
  for details).}
\tablenotetext{c}{$B$, $R$, and $I$ magnitudes are photographic magnitudes
  from the USNO-B1 catalogue.  The $B$ and $R$ magnitudes are preferentially
  from the second epoch plates; if no second epoch magnitude was available,
  then the first epoch magnitude was used.}
\tablenotetext{d}{$J$, $H$, and $K_s$ magnitudes are from the 2MASS
  final release point source catalogue \citep{2mass}.}
\tablenotetext{e}{Number of surveys out of the 5 in USNO-B1 on which
  the object was detected in the construction of USNO-B1.} 
\tablenotetext{f}{Classification: d = dwarf, sd = sub-dwarf, wd =
  white dwarf.}
\tablenotetext{g}{Alternate Identification, if this is a previously known
  object.  TYC2=\citet{hetal00}, NLTT=\citet{nltt}, LHS=\citet{lhs},
  LDS=\citet{lds}, UCAC2=\citet{ucac2}.}
\tablenotetext{h}{First epoch magnitude.}
\end{deluxetable}

\clearpage

\begin{deluxetable}{lrrrrrrrrrrrrrcl}
\rotate
\tabletypesize{\scriptsize}
\tablecaption{Serendipitous objects with new Position and Proper Motion Solutions. \label{tbl-sersel}}
\tablewidth{0pt}
\tablehead{
\colhead{ID\tablenotemark{a}} & 
\colhead{$\mu$\tablenotemark{b}} &
\colhead{$\theta$\tablenotemark{b}} & 
\colhead{RA} & 
\colhead{Dec} &
\colhead{{\it l}} &
\colhead{{\it b}} &
\colhead{$\rm B$\tablenotemark{c}} & 
\colhead{$\rm R$\tablenotemark{c}} & 
\colhead{$\rm I$\tablenotemark{c}} & 
\colhead{$\rm J$\tablenotemark{d}} & 
\colhead{$\rm H$\tablenotemark{d}} & 
\colhead{$\rm K_s$\tablenotemark{d}} & 
\colhead{\#\tablenotemark{e}} &
\colhead{Class\tablenotemark{f}} &
\colhead{AltID\tablenotemark{g}} \\
\colhead{} & 
\colhead{[${\rm mas \, yr^{-1}}$]} &
\colhead{[deg]} & 
\colhead{[hrs]} & 
\colhead{[deg]} &
\colhead{[deg]} &
\colhead{[deg]} &
\colhead{[mag]} & 
\colhead{[mag]} & 
\colhead{[mag]} & 
\colhead{[mag]} & 
\colhead{[mag]} & 
\colhead{[mag]} &
\colhead{} & 
\colhead{} &
\colhead{} \\
\colhead{(1)} & 
\colhead{(2)} &
\colhead{(3)} & 
\colhead{(4)} & 
\colhead{(5)} &
\colhead{(6)} &
\colhead{(7)} & 
\colhead{(8)} &
\colhead{(9)} &
\colhead{(10)} &
\colhead{(11)} &
\colhead{(12)} & 
\colhead{(13)} & 
\colhead{(14)} & 
\colhead{(15)} &
\colhead{(16)}
}
\startdata
MUSR 01  &  155.5 &   94.3 &  0.30014 &  76.9962 & 120.9982 &  14.2501 & 18.2 & 16.3 & 14.8 & 13.0 & 12.5 & 12.2 &  3 & d &  \\
MUSR 04  &   64.2 &   39.9 &  0.36642 &  76.2059 & 121.1315 &  13.4372 & 17.5 & 15.9 & 14.6 & 13.0 & 12.4 & 12.1 &  3 & d &  \\
MUSR 07  &   91.4 &   95.6 &  0.52378 &  77.4410 & 121.8091 &  14.6129 & 19.3 & 17.2 & 15.7 & 14.2 & 13.5 & 13.3 &  4 & d &  \\
MUSR 13  &  114.0 &  121.7 &  1.86147 &  47.4891 & 133.3661 & -14.1583 & 15.8 & 15.0 & 11.4 & 10.7 & 10.1 &  9.8 &  3 & d &  \\
MUSR 16  &  137.4 &  127.3 &  3.74106 &  78.0435 & 131.5250 &  18.1683 & 19.1 & 16.9 & 15.4 & 14.1 & 13.5 & 13.3 &  3 & d & UB-1680-0031453 \\
MUSR 17  &  194.3 &  122.7 &  3.84409 &  52.2317 & 148.5094 &  -1.5079 & 17.7 & 16.0 & 14.2 & 12.7 & 12.1 & 11.8 &  3 & d & UB-1422-0127233 \\
MUSR 19  &   85.3 &  130.6 &  4.26547 &  63.3473 & 143.6395 &   9.0300 & 16.8\tablenotemark{i} & 14.8\tablenotemark{i} & 13.4\tablenotemark{i} & 12.7 & 12.0 & 11.8 &  2 & d &  \\
MUSR 21  &  185.9 &  120.6 &  4.44636 &  63.1542 & 144.6443 &   9.7688 & 18.6 & 16.3 & 14.5 & 13.3 & 12.8 & 12.5 &  3 & d & NLTT-13207 \\
MUSR 22  &   85.1 &  138.6 &  5.22302 &  52.9614 & 156.5483 &   8.1374 & 15.1 & 13.8 & 13.4 & 12.2 & 11.5 & 11.4 &  3 & d &  \\
MUSR 23  &  150.3 &  115.9 &  5.25623 &  67.5593 & 144.2512 &  16.4504 & 18.5 & 16.6 & 14.9 & 13.7 & 13.2 & 13.0 &  3 & d &  \\
MUSR 25  &   41.7 &  210.3 &  5.52282 &  67.4864 & 145.1243 &  17.7331 & 16.6 & 15.4\tablenotemark{h} & \nodata & 13.5 & 12.9 & 12.8 &  3 & d & UB-1574-0111219 \\
MUSR 27  &  230.9 &  145.9 &  5.84567 &  62.9571 & 150.3206 &  17.4703 & 13.5 & 12.4\tablenotemark{{\rm h,i}} & 10.7\tablenotemark{i} &  9.9 &  9.3 &  9.2 &  3 & d &  \\
MUSR 31  &   57.6 &   74.8 & 10.28827 &  67.6570 & 141.9380 &  43.3458 & 17.3 & 15.4 & 14.1 & 13.3 & 12.6 & 12.4 &  4 & d &  \\
MUSR 35  &  115.7 &  286.0 & 11.95683 &  -3.2109 & 277.5956 &  56.9792 & 19.6 & 17.0 & 15.4 & 14.1 & 13.5 & 13.2 &  4 & d &  \\
MUSR 39  &  156.7 &  219.6 & 12.71622 &  65.7056 & 124.3272 &  51.3997 & 17.5 & 16.8\tablenotemark{h} & 15.7\tablenotemark{i} & 16.6 & 17.1 & 16.1 &  3 & wd & UB-1557-0144114 \\
MUSR 40  &  150.7 &  187.5 & 15.06479 &  68.2484 & 106.4694 &  44.4086 & 18.6\tablenotemark{h} & 16.3\tablenotemark{h} & 15.4 & 14.7 & 14.3 & 14.0 &  3 & sd & UB-1582-0176403 \\
MUSR 42  &  101.2 &  181.9 & 15.67851 &  68.6857 & 103.8286 &  41.6052 & 19.2\tablenotemark{h} & 16.5\tablenotemark{h} & \nodata & 14.4 & 13.8 & 13.6 &  2 & d &  \\
MUSR 44  &  138.0 &  210.5 & 18.18237 &   6.3899 &  34.0677 &  11.9007 & 17.3 & 15.2 & 14.6 & 13.3 & 12.6 & 12.5 &  3 & d &  \\
MUSR 45  &  256.0 &  259.9 & 20.13456 &  62.8713 &  96.3361 &  15.8208 & 18.8\tablenotemark{h} & 16.5 & 15.0 & 13.4 & 12.9 & 12.7 &  4 & d &  \\
MUSR 46  &  102.6 &   38.1 & 21.28918 &  61.1952 &  99.8812 &   8.3154 & 18.1 & 15.8 & 14.1 & 12.6 & 11.9 & 11.7 &  3 & d &  \\
MUSR 49  &   49.6 &   90.3 & 22.38069 &  43.0352 &  96.1106 & -11.9656 & 17.9 & 15.3 & 13.6 & 13.4 & 12.7 & 12.5 &  4 & d &  \\
MUSR 51  &   89.7 &   52.7 & 22.52580 &  60.3778 & 106.4630 &   2.0585 & 18.1 & 15.9 & 14.8 & 13.5 & 12.9 & 12.6 &  3 & d & UB-1503-0343496 \\
MUSR 52  &  119.0 &   78.1 & 22.54173 &  64.3235 & 108.5745 &   5.3966 & 19.3\tablenotemark{h} & 16.4\tablenotemark{h} & \nodata & 13.5 & 12.9 & 12.7 &  2 & d &  \\
MUSR 53  &   89.2 &  119.2 & 22.62082 &  63.5496 & 108.6377 &   4.4657 & 18.3\tablenotemark{h} & 16.2\tablenotemark{h} & \nodata & 13.1 & 12.4 & 12.2 &  2 & d &  \\
MUSR 54  &  144.0 &   14.8 & 22.64093 &  62.5719 & 108.2758 &   3.5469 & 16.4 & 14.5 & 14.1 & 13.1 & 12.5 & 12.4 &  3 & sd &  \\
MUSR 55  &   44.8 &  233.1 & 22.66454 &  66.7672 & 110.4644 &   7.1352 & 17.9 & 15.8 & 15.3 & 14.0 & 13.2 & 13.0 &  5 & d &  \\
MUSR 56  &  127.4 &  107.4 & 22.66674 &  64.3870 & 109.3132 &   5.0483 & 14.3 & 12.2 & 10.6 &  9.6 &  9.0 &  8.8 &  3 & d & LDS-4990 \\
MUSR 61  &  129.6 &   71.9 & 22.76775 &  64.5190 & 109.9534 &   4.8543 & 16.7 & 14.9 & 13.8 & 12.2 & 11.6 & 11.4 &  3 & d &  \\
MUSR 62  &  149.7 &   62.5 & 22.76936 &  64.5065 & 109.9568 &   4.8384 & 15.2\tablenotemark{h} & 14.1\tablenotemark{h} & \nodata & 11.5 & 10.9 & 10.8 &  3 & d &  \\
MUSR 63  &  122.0 &  100.0 & 22.76986 &  63.6424 & 109.5576 &   4.0713 & 19.8\tablenotemark{h} & 19.3 & 17.9 & 13.6 & 12.9 & 12.6 &  4 & d &  \\
MUSR 64  &  107.7 &  263.0 & 22.79564 &  65.6343 & 110.6280 &   5.7618 & 15.5\tablenotemark{i} & 13.5\tablenotemark{i} & 13.1\tablenotemark{i} & 11.7 & 11.1 & 11.0 &  3 & d &  \\
MUSR 65  &   95.8 &   44.1 & 22.83035 &  64.2553 & 110.1941 &   4.4338 & 17.8\tablenotemark{h} & 16.4\tablenotemark{h} & \nodata & 14.4 & 13.8 & 13.7 &  2 & sd &  \\
MUSR 67  &  105.4 &  241.7 & 22.94335 &  66.5543 & 111.8480 &   6.1862 & 20.3\tablenotemark{{\rm h,i}} & 17.1 & 15.4 & 13.7 & 13.1 & 12.8 &  3 & d &  \\
MUSR 76  &   85.9 &   83.7 & 23.68831 &  78.5560 & 119.3635 &  16.1732 & 19.8\tablenotemark{h} & 16.6\tablenotemark{h} & \nodata & 14.2 & 13.6 & 13.4 &  2 & d &  \\
MUSR 77  &   80.6 &   83.2 & 23.72320 &  78.5201 & 119.4573 &  16.1100 & 19.9\tablenotemark{h} & 16.8\tablenotemark{h} & \nodata & 14.2 & 13.6 & 13.3 &  2 & d &  \\
MUSR 78  &  125.8 &   82.8 & 23.73681 &  79.7080 & 119.8311 &  17.2433 & 18.8\tablenotemark{h} & 16.3\tablenotemark{h} & \nodata & 13.7 & 13.1 & 12.9 &  2 & d &  \\
MUSR 80  &   97.9 &   55.7 & 23.83274 &  78.5106 & 119.7836 &  16.0163 & 16.2 & 14.6 & 12.8 & 11.0 & 10.3 & 10.1 &  3 & d &  \\
MUSR 81  &   86.9 &   83.3 & 23.94682 &  79.1474 & 120.2736 &  16.5586 & 17.0 & 15.2 & 13.6 & 12.0 & 11.5 & 11.2 &  4 & d &  \\
MUSR 82  &  138.3 &   65.3 & 23.97094 &  78.9788 & 120.3055 &  16.3792 & 15.4\tablenotemark{i} & 14.4 & 13.8 & 13.1 & 12.7 & 12.6 &  3 & sd &  \\
\enddata
\tablenotetext{a}{IDs of the form ZZZZ-NNNNNNN are from USNO-B1, and those
  of the form ZZZZ-RRRRR-N are from Tycho-2.}
\tablenotetext{b}{The proper motions are relative to the reference
  frame established by the YS4.0 catalogue stars (see Monet et al. 2003
  for details).}
\tablenotetext{c}{$B$, $R$, and $I$ magnitudes are photographic magnitudes
  from the USNO-B1 catalogue.  The $B$ and $R$ magnitudes are preferentially
  from the second epoch plates; if no second epoch magnitude was available,
  then the first epoch magnitude was used.}
\tablenotetext{d}{$J$, $H$, and $K_s$ magnitudes are from the 2MASS
  final release point source catalogue \citep{2mass}}
\tablenotetext{e}{Number of surveys out of the 5 in USNO-B1 on which
  the object was detected in the construction of USNO-B1.} 
\tablenotetext{f}{Classification: d = dwarf, sd = sub-dwarf, wd =
  white dwarf.}
\tablenotetext{g}{Alternate Identification, if this is a previously known
  object. NLTT=\citet{nltt}, LDS=\citet{lds}, UB=USNO-B1 object ID
  where the USNO-B1 has a decent position and motion match, but used
  incomplete data.}
\tablenotetext{h}{First epoch magnitude.}
\tablenotetext{i}{Magnitude taken from another USNO-B1 catalogue entry,
  which was made up of additional detections of this objects.}
\end{deluxetable}

\clearpage

\clearpage

\clearpage

\clearpage

\clearpage

\clearpage

\clearpage

\clearpage

\clearpage

\clearpage

\clearpage

\clearpage

\clearpage

\clearpage

\clearpage

\clearpage

\clearpage







\begin{thebibliography}{}

\bibitem[Abazajian et al.(2003)]{aetal03} Abazajian, K., et al. 2003, \aj,
  126, 2081(SDSS DR1)

\bibitem[Bakos et al.(2002)]{bsn02} Bakos, G. A., Sahu,
  K. C., \& Nemeth, P. 2002, \apjs, 141,187 (rLHS; Vizier Catalogue:
  I/279)

\bibitem[Cutri et al.(2003)]{2mass} Cutri, R. M. et al. 2003,
  Explanatory Supplement to the 2MASS All Sky Data Release (Pasadena:
  IPAC)(http://www.ipac.caltech.edu/2mass/releases/allsky/doc/explsup.html)


\bibitem[Deacon et al.(2005)]{dhc05} Deacon, N. R., Hambly, N. C.,
  \& Cooke, J. A. 2005, \aap, submitted (astro-ph/0412127)

\bibitem[Dehnen \& Binney(1998)]{db98} Dehnen, W., \& Binney, J. 1998,
  \mnras, 298, 387

\bibitem[Delfosse et al.(2001)]{detal01} Delfosse, X., et al. 2001, \aap,
  366, L13

\bibitem[ESA(1997)]{hip} European Space Agency 1997, The Hipparcos and
  Tycho Catalogues (SP-1200; Noordwijk: ESA) (Vizier Catalogue:
  I/239)

\bibitem[Giclas et al.(1971)]{gic71} Giclas, H. L.,
  Burnham Jr., R, \& Thomas, N. G. 1971, Lowell Proper Motion Survey:
  Northern Hemisphere, (Flagstaff: Lowell Observatory) (Vizier Catalogue:
  I/79)

\bibitem[Giclas et al.(1978)]{gic78} Giclas, H. L.,
  Burnham Jr., R, \& Thomas, N. G. 1978, Lowell Proper Motion Survey:
  Southern Hemisphere, (Flagstaff: Lowell Observatory) (Vizier Catalogue:
  I/112)

\bibitem[Gould(2003)]{g03} Gould, A. 2003, \aj, 126, 472

\bibitem[Gould \& Kollmeier(2004)]{gk04} Gould, A., \& Kollmeier,
  J. A., 2004, \apjs, 152, 103

\bibitem[Gould \& Salim(2003)]{gs03} Gould, A., \& Salim, S. 2003,
  \apj, 582, 1001

\bibitem[Hambly et al.(2004)]{hetal04} Hambly, N. C., Henry, T., J.,
  Subsavage, J. P., Brown, M. A., \& Jao, W.-C. 2004, \aj, 128, 437

\bibitem[H{\o}g et al.(2000)]{hetal00} H{\o}g, E., et al. 2000, \aap,
  355, L27 (Vizier Catalogue: I/259)

\bibitem[L\'epine(2005)]{l05}  L\'epine, S. 2005, \aj, submitted
  (astro-ph/0501266) 

\bibitem[L\'epine et al.(2002)]{lsr02}  L\'epine, S., Shara,
  M. M., \& Rich, R. M. 2002, \aj, 124, 1190 (LSR)

\bibitem[L\'epine et al.(2003a)]{lrs03a}  L\'epine, S., Rich, R. M., \&
  Shara, M. M. 2003, \aj, 125, 1598

\bibitem[L\'epine et al.(2003b)]{lsr03b}  L\'epine, S., Shara, M. M., \&
   Rich, R. M. 2003, \aj, 126, 921

\bibitem[Luyten(1940)]{lds} Luyten, W. J. 1940, The LDS Catalogue,
  Publ. of the Astron. Obs. Univ. Minnesota vol. III part 3,
  (Minneapolis, University of Minnesota) (Vizier Catalogue: I/130)

\bibitem[Luyten(1979a)]{lhs} Luyten, W. J. 1979, LHS Catalog,
  (Minneapolis, University of Minnesota) (Vizier Catalogue: I/87B)

\bibitem[Luyten(1979b)]{nltt} Luyten, W. J. 1979, NLTT Catalog,
  (Minneapolis, University of Minnesota) (Vizier Catalogue: I/98A)


\bibitem[Monet et al.(2000)]{metal00} Monet, D. G., Fisher, M. D.,
  Liebert, J., Canzian, B., Harris, H. C., \& Reid, I. N. 2000, \aj,
  120, 1541

\bibitem[Monet et al.(1996)]{metal96} Monet, D. G. et al. 1996,
  USNO-A1.0 (Flagstaff: U. S. Naval Obs.) (Vizier Catalogue: I/243)

\bibitem[Monet et al.(1998)]{metal98} Monet, D. G. et al. 1998,
  USNO-A2.0 (Flagstaff: U. S. Naval Obs.) (Vizier Catalogue: I/252)

\bibitem[Monet et al.(2003)]{metal03} Monet, D. G., et al. 2003, \aj,
  125, 984 (USNO-B1, Vizier Catalogue: I/284)

\bibitem[Munn et al.(2004)]{metal04} Munn, J. A., et al. 2004, \aj, 127, 3034


\bibitem[Olling \& Dehnen(2003)]{od03} Olling, R. P., \& Dehnen,
  W. 2003, \apj, 599, 275

\bibitem[Oppenheimer et al.(2001)]{oetal01} Oppenheimer, B. R., Hambly,
  N. C., Digby, A. P., Hodgkin, S. T., \& Saumon, D. 2001, Science,
  292, 5517

\bibitem[Pokorny et al.(2003)]{petal03} Pokorny, R. S., Jones, H. R. A.,
  \& Hambly, N. C. 2003, \aap, 387, 575

\bibitem[Pokorny et al.(2004)]{petal04} Pokorny, R. S., Jones, H. R. A.,
  Hambly, N. C., \& Pinfield, D. J. 2004, \aap, 421, 763

\bibitem[Reyl\'e \& Robin(2004)]{rr04} Reyl\'e, C., \& Robin, A. C. 2004, 
  \aap, 421, 643

\bibitem[Reyl\'e et al.(2002)]{rrsi02} Reyl\'e, C., Robin, A. C.,
  Scholz, R. D., \& Irwin, M. 2002, \aap, 390, 491

\bibitem[Salim \& Gould(2002)]{sg02} Salim, S., \& Gould, A. 2002,
  \apj, 575, L83

\bibitem[Salim \& Gould(2003)]{sg03} Salim, S., \& Gould, A. 2003,
  \apj, 582, 1011

\bibitem[van Biesbroeck(1961)]{vb61} van Biesbroeck, G. 1961, \aj, 66,
  528

\bibitem[Voges et al.(1999)]{rxs} Voges, W. et al. 1999, \aap, 349,
  389 (1RXS, Vizier Catalogue: IX/10A)

\bibitem[Zacharias et al.(2004)]{ucac2} Zacharias, N. et al. 2004, \aj, 
  127, 3043 (Vizier Catalogue: I/289)

\end{thebibliography}
\end{document}